\documentclass[english,notitlepage,nofootinbib]{revtex4-2}
\usepackage[T1]{fontenc}
\usepackage[latin9]{inputenc}
\usepackage{color}
\usepackage{babel}
\usepackage{amsmath}
\usepackage{amssymb}
\usepackage{graphicx}
\usepackage[bookmarks=false,
 breaklinks=false,pdfborder={0 0 1},backref=false,colorlinks=false]
 {hyperref}

\makeatletter

\providecommand{\tabularnewline}{\\}

\usepackage{babel}

\usepackage{xcolor}
\usepackage{pdfcolmk}

\usepackage{ulem}

\providecolor{lyxadded}{rgb}{0,0,1}
\providecolor{lyxdeleted}{rgb}{1,0,0}

\DeclareRobustCommand{\lyxsout}[1]{\ifx\\#1\else\sout{#1}\fi}
\allowdisplaybreaks

\makeatother

\begin{document}
\title{Inclusive hadroproduction of charmonia-bottomonia pairs in the CGC
framework}
\author{Marat Siddikov}
\affiliation{Departamento de Física, Universidad Técnica Federico Santa
María,~~~~\\
 y Centro Científico - Tecnológico de Valparaíso, Casilla 110-V, Valparaíso,
Chile}
\begin{abstract}
In this preprint we analyze the inclusive hadroproduction of heavy
charmonia-bottomonia pairs in the Color Glass Condensate framework
in the dilute-dense approximation. The production mechanisms can be
classified by number of gluons emitted from projectile as Single-
and Double Parton Scattering. We analyzed separately both types of
contributions and evaluated the corresponding impact factors in the
leading order in strong coupling $\alpha_{s}$. For the Double Parton
Scattering the impact factors factorize into a product of impact factors
of single charmonia and bottomonia production cross-sections, and
the cross-section in the large-$N_{c}$ limit reduces to the well-known
pocket formula. For the Single Parton Scattering we found that the
cross-section of the process is sensitive to dipole, quadrupole, sextupole
and octupole scattering amplitudes (correlators of Wilson lines),
thus opening possibility to study phenomenologically these novel objects.
We also made phenomenological estimates of the cross-sections in the
dilute approximation and found that the suggested approach gives
a fair qualitative description of the LHCb data for $J/\psi+\Upsilon(1S)$
production.

\end{abstract}
\pacs{12.38.-t, 14.40.Pq 13.60.Le}
\keywords{Quantum chromodynamics, Heavy quarkonia, Meson production}
\maketitle

\section{Introduction}

The quarkonia have been considered as useful probes of the gluonic
field of the target almost since their discovery. The small sizes
of the quarkonia and the heavy masses of the constituent quarks, which
play the role of the natural hard scales, motivated application of
various perturbative schemes for description of their production.
The hadronization of heavy quarks into quarkonia at present is understood
reasonably well and is described in the Nonrelativistic QCD (NRQCD)
framework, which encodes it in terms of the nonperturbative Long Distance
Matrix Elements (LDMEs) of quarkonia states~~\cite{Bodwin:1994jh,Maltoni:1997pt,Brambilla:2008zg,Feng:2015cba,Brambilla:2010cs}.
This approach allows to get a reasonable estimates for the cross-sections,
and for this reason (single) quarkonia production is being included
in phenomenological fits of the gluonic generalized parton distributions
and color dipole cross-sections~\cite{Kowalski:2006hc,RESH,Rezaeian:2012ji}
which control the cross-section in this channel in different kinematics.
From this point of view, the production of multiple quarkonia, especially
the heavy quarkonia pairs presents an interesting tool which can be
used to improve our understanding of the heavy quarkonia formation.
Such channels attracted the theoretical attention since the early
days of QCD~\cite{Brodsky:1986ds,Lepage:1980fj,Berger:1986ii,Baek:1994kj},
and the possibility to study such channels experimentally was demonstrated
soon afterwards~\cite{NA3:1,NA3:2}. Recent experiments at LHC demonstrated
a possibility to study the production of quarkonia pairs with reasonably
high precision. 

In this paper we will focus on the associated production of charmonia
and bottomonia pairs, which has a very simple structure of the partonic
amplitude. The theoretical studies of such processes (particularly,
$J/\psi+\Upsilon$ hadroproduction) have been initiated in~\cite{Ko:2010xy,Likhoded:2015zna,Shao:2016wor}
and attracted a lot of theoretical attention. In the collinear and
$k_{T}$ factorization approach, it is known that the contribution
of SPS in such processes is strongly suppressed by a strong coupling
$\alpha_{s}^{2}$ compared to production of quarkonia pairs with the
same hidden flavor. For this reason, the suggested processes have
been suggested as a possible tool for studies of various mechanisms
that are usually suppressed and controlled by the poorly known color
octet Long Distance Matrix Elements or the double parton distribution
functions (DPDFs). The recent experimental data~\cite{LHCb:2023qgu,D0:2015dyx}
demonstrated that the contribution controlled by DPDFs indeed may
be significant in this channel. However, at present the situation
remains unclear: while inclusion of the DPDFs allows to explain the
difference between the data and predictions based on single-parton
scattering, the value of the so-called parameter $\sigma_{{\rm eff}}$,
which controls the magnitude of the DPDFs, depends significantly on
the channel used for its extraction~\cite{Belyaev:2017sws,Lansberg:2016rcx,Lansberg:2019adr}.
 This discrepancy hints that this channel in the small-$x$ kinematics
potentially can get contributions from other mechanisms, which eventually
lead to onset of saturation effects, and for this reason it is appropriate
to analyze this process in the frameworks with built-in saturation.
In what follows we will realize such study using the Color Glass Condensate
(CGC) framework~~ \cite{McLerran:1993ni,McLerran:1993ka,McLerran:1994vd,Gelis:2010nm,Iancu:2003uh},
which naturally incorporates the saturation effects and provides a
phenomenologically reasonable description of both hadron-hadron and
lepton-hadron collisions~\cite{Aidala:2020mzt,Ma:2014mri,Cheung:2024qvw,Mantysaari:2020lhf,Kang:2023doo,Tuchin:2004,Blaizot:2004,ALICE:2012,ATLAS:2016,IPSat,watt:bcgc,DaSilveira:2018haa,Albacete:2012td,LHCb:2022ahs,Mantysaari:2023xcu}.
The cross-sections of physical processes in this framework are expressed
in terms of the forward multipole scattering amplitudes ($n$-point
correlators of Wilson lines), which present important physical characteristics
of the target. In some special cases these correlators may be related
to the forward dipole scattering amplitudes, which are well-known
from the phenomenology. 

The paper is structured as follows. Below in Section~\ref{sec:FrameworkDerivation}
we briefly describe the main components of the CGC framework and then
present the theoretical results for the hadroproduction of heavy quarkonia
pairs in this approach. In Section~\ref{sec:Numer} we provide
numerical estimates using the phenomenological parametrizations of
dipole, quadrupole, sextupole and octupole amplitudes. Finally, in
Section~\ref{sec:Conclusions} we draw conclusions.

\section{Theoretical framework}

\label{sec:FrameworkDerivation}In what follows we will analyze the
hadroproduction of the quarkonia pairs in a dilute-dense limit. Such
treatment is justified in the forward kinematics, when the heavy partons
and quarkonia move with relatively small rapidities with respect to
one of the colliding protons (``projectile''). The interaction of
the latter with the partons may be described in partonic picture,
disregarding possible saturation effects. At the same time, the partons
move with very large energies with respect to another hadron (``target''),
and thus a CGC framework should be applied in order to describe this
interaction and onset of the saturation effects. 

\begin{figure}
\includegraphics[width=8cm]{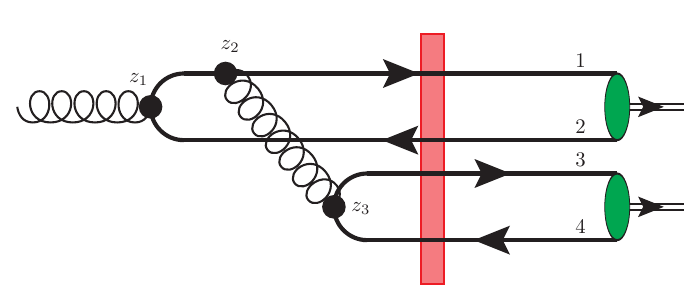}\includegraphics[width=8cm]{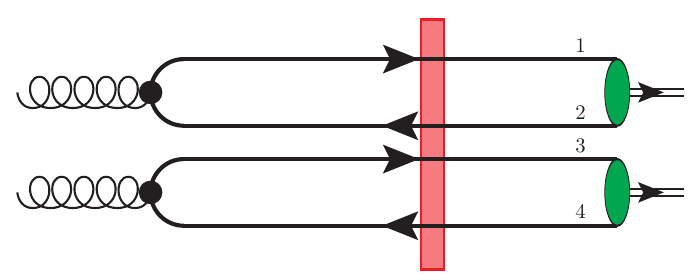}

\caption{\protect\label{fig:CGCBasic-2}Left and right plot: Example diagrams
which describe the inclusive heavy quarkonia pair production in single-
and double parton scattering (SPS and DPS respectively). The red block
represents a shock wave in the CGC picture. }
\end{figure}

The production of the heavy quarkonia pair in this picture may be
considered as a result of the scattering of one or two gluons emitted
from the projectile in the field of the target. In what follows we
will refer to these processes as Single- and Double Parton Scattering
(SPS and DPS respectively, see Figure~\ref{fig:CGCBasic-2} for illustration),
in agreement with the terminology used in the analyses based on the
collinear and $k_{T}$-factorization (see~\cite{Lansberg:2019adr}
for a review). In the following Subsection~\ref{subsec:Kinematics}
we introduce our notations and the light-cone decomposition of the
particle's momenta. In Subsection~~\ref{subsec:Derivation} we briefly
introduce the CGC approach which will be used for theoretical evaluations. Finally, in Subsections
\ref{subsec:SPS} and \ref{subsec:DPS} we derive the expressions
for the contributions of SPS and DPS mechanisms. 

\subsection{Kinematics of the process}

\label{subsec:Kinematics}In what follows we will work in the gluon-proton
scattering frame, assuming that the momentum of the target and the
total momentum of the gluons(s) emitted from projectile point in direction
of axis $z$. We will use a notation $q$ for the momentum of the
incoming gluon in SPS mechanism, $\mathfrak{q}_{1},\,\,\mathfrak{q}_{2}$
for the momenta of incoming (projectile) gluons in DPS mechanism,
$P$ for the target momentum, and $p_{1},p_{2}$ for the momenta of
the produced quarkonia. For the sake of definiteness, we will use
the Kogut-Soper convention~\cite{Brodsky:1997de} for the light-cone
components of all the vectors, and bold-face letters with subindex
$\perp$ for the transverse (1,2)-components, namely
\begin{equation}
v^{\mu}=\left(v^{+},\quad v^{-},\quad\boldsymbol{v}_{\perp}\right),\qquad v^{\pm}=\frac{v^{0}\pm v^{3}}{\sqrt{2}},\qquad\boldsymbol{v}_{\perp}=v_{x}\hat{\boldsymbol{x}}+v_{y}\hat{\boldsymbol{y}}.
\end{equation}
The square of the vector in this convention is given by 
\begin{equation}
v^{2}\equiv v^{\mu}v_{\mu}=2v^{+}v^{-}-\boldsymbol{v}_{\perp}^{2},
\end{equation}
and the convolution with Dirac $\gamma$-matrix reads as 
\begin{equation}
\hat{v}=\gamma^{\mu}v_{\mu}=\gamma^{+}v^{-}+\gamma^{-}v^{+}-\boldsymbol{\gamma}_{\perp}\cdot\boldsymbol{v}_{\perp}.\label{eq:gammaSlash}
\end{equation}
The light-cone decomposition of the momentum $P$ of the target (proton)
and the momenta $p_{1},p_{2}$ of the produced quarkonia are given
by

\begin{align}
P & =\left(\frac{m_{N}^{2}}{2P^{-}},\,P^{-},\,\,\boldsymbol{0}_{\perp}\right),\quad P^{-}=\frac{E_{p}+\sqrt{E_{p}^{2}-m_{N}^{2}}}{\sqrt{2}}\approx\sqrt{2}E_{p}
\end{align}
\begin{align}
p_{a} & =\left(M_{a,\perp}\,e^{y_{a}}\,,\,\frac{M_{a,\perp}e^{-y_{a}}}{2},\,\,\boldsymbol{p}_{a}^{\perp}\right),\quad M_{a,\perp}\equiv\sqrt{M_{a}^{2}+\left(\boldsymbol{p}_{a}^{\perp}\right)^{2}},\quad a=1,2,\label{eq:MesonLC}
\end{align}
where $\left(y_{a},\boldsymbol{p}_{a}^{\perp}\right)$ are the rapidities
and transverse momenta of the produced quarkonia with respect to collision
axis, $M_{a}$ are their masses, and $m_{N}$ is the mass of the nucleon.
In this manuscript we will be mostly interested in the high-energy
collider kinematics, when quarkonia are produced with relatively small
transverse momenta. In this kinematics it is possible to use eikonal
picture and assume that the plus-component $q_{a}^{+}$ of the incoming
gluon light-cone momentum is shared only between the produced quarkonia.
This allows us to write out the light-cone decomposition of the incoming
photon momentum in SPS process as 
\begin{align}
q & =\,\left(q^{+},\,0,\,\,\boldsymbol{0}_{\perp}\right),\qquad q^{+}=M_{1,\perp}\,e^{y_{1}}+M_{2,\perp}\,e^{y_{2}}.
\end{align}
Similarly, for a pair of two incoming gluons in the DPS kinematics
we may assume that the momenta of the gluons emitted from the projectile
are given by
\begin{align}
\mathfrak{q}_{a} & =\,\left(M_{a,\perp}\,e^{y_{a}},\,0,\,\,\boldsymbol{\mathfrak{q}}_{a,\perp}\right),\quad a=1,2.
\end{align}
For the sake of brevity, sometimes we will use notations
\[
\alpha_{1}=\frac{M_{1,\perp}\,e^{y_{1}}}{M_{1,\perp}\,e^{y_{1}}+M_{2,\perp}\,e^{y_{2}}},\quad\alpha_{2}=1-\alpha_{1}=\frac{M_{2,\perp}\,e^{y_{2}}}{M_{1,\perp}\,e^{y_{1}}+M_{2,\perp}\,e^{y_{2}}}
\]
which allow us to rewrite the momenta of the quarkonia in the compact
form as
\[
p_{a}=\left(\alpha_{a}q^{+}\,,\,\frac{M_{a,\perp}^{2}}{2\alpha_{a}q^{+}},\,\,\boldsymbol{p}_{a}^{\perp}\right),\quad a=1,2,
\]
\begin{align}
q_{({\rm SPS})} & =\,\left(q^{+},\,0,\,\,\boldsymbol{0}_{\perp}\right),\qquad\mathfrak{q}_{a}=\,\left(\alpha_{a}q^{+},\,0,\,\,\boldsymbol{\mathfrak{q}}_{a,\perp}\right).
\end{align}
 The invariant energy $W$ of the gluon-proton collision and the invariant
mass $M_{12}$ of the produced heavy quarkonia pair may be expressed
as 
\begin{equation}
W^{2}\equiv\left(q+P\right)^{2}=-Q^{2}+m_{N}^{2}+2q\cdot P\approx-m_{N}^{2}+2P^{-}\left(M_{1}^{\perp}\,e^{y_{1}}+M_{2}^{\perp}\,e^{y_{2}}\right),\label{eq:W2}
\end{equation}
and 
\begin{equation}
{\mathcal{M}}_{12}^{2}=\left(p_{1}+p_{2}\right)^{2}=M_{1}^{2}+M_{2}^{2}+2\left(M_{1}^{\perp}M_{2}^{\perp}\cosh\Delta y-\boldsymbol{p}_{1}^{\perp}\cdot\boldsymbol{p}_{2}^{\perp}\right)\label{eq:M12}
\end{equation}
respectively. In order to avoid possible (soft) final state interactions
between quarkonia, in what follows we will tacitly assume that these
heavy mesons are kinematically well-separated from each other, namely
that the invariant mass ${\mathcal{M}}_{12}^{2}$ is sufficiently
large, thus avoiding a near-threshold production.

\subsection{High energy scattering in the CGC picture}

\label{subsec:Derivation} In the CGC framework, which is valid in the
small-$x$ (high energy) kinematics, the interaction of the heavy
parton with the target in configuration space is described in eikonal
approximation, and is given by the Wilson line $U(\boldsymbol{x}_{\perp})$~\cite{McLerran:1993ni,McLerran:1993ka,McLerran:1994vd,MUQI,MV,Gelis:2010nm,Iancu:2003uh}
\begin{equation}
U\left(\boldsymbol{x}_{\perp}\right)=P\exp\left(ig\int dx^{-}A_{a}^{+}\left(x^{-},\,\boldsymbol{x}_{\perp}\right)t^{a}\right),\label{eq:Wilson}
\end{equation}
where $\boldsymbol{x}_{\perp}$ is the transverse coordinate (impact
parameter) of the parton, and $t_{a}$ are the generators of the color
group generators in the corresponding representation ($\boldsymbol{3}$,
$\boldsymbol{\bar{3}}$ or $\boldsymbol{8}$ for quark, antiquark
and gluon, respectively), $A_{\mu}^{a}(x)$ is the semiclassical gluonic
field created color charge distributed with density $\rho_{a}(\boldsymbol{x})$,
with $A_{\mu}^{a}(x)=-\frac{1}{\nabla_{\perp}^{2}}\rho_{a}(x^{-},\,\boldsymbol{x}_{\perp})$
. In the momentum space the interactions of the partons with the target
can be rewritten as CGC Feynman rules~~\cite{Blaizot:2004,Ayala:2017rmh,Caucal:2021ent,Caucal:2022ulg}
provided in Table~\ref{tab:FR}. The prefactor $2\pi\delta\left(\ell^{+}-\ell'{}^{+}\right)$
in all the shockwave interaction vertices reflects conservation of
the plus-component and sometimes is rewritten in the literature as
\begin{equation}
2\pi\delta\left(\ell^{+}-\ell'{}^{+}\right)=\int_{-\infty}^{+\infty}dz^{-}e^{i\left(\ell^{+}-\ell^{'+}\right)z^{-}}
\end{equation}

\begin{table}
\begin{tabular}{|c|c|}
\hline 
\begin{minipage}[t]{0.4\columnwidth}%
\textcolor{white}{.}\\
 \includegraphics[scale=0.7]{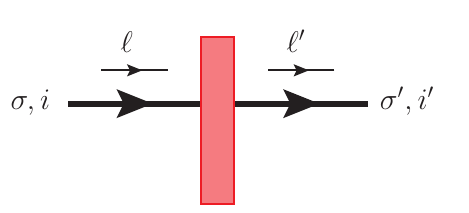}%
\end{minipage} & %
\begin{minipage}[t]{0.55\columnwidth}%
The interaction vertex of the quark with the shock wave: 
\[
T_{\sigma,\sigma';\,i,i'}^{Q}=2\pi\delta\left(\ell^{+}-\ell'{}^{+}\right)\gamma_{\sigma',\,\sigma}^{+}\int d^{2}\boldsymbol{z}e^{i\left(\boldsymbol{\ell}-\boldsymbol{\ell}'\right)\cdot\boldsymbol{z}}\left[U\left(\boldsymbol{z}\right)-1\right]_{i',i}
\]
\end{minipage}\tabularnewline
\hline 
\begin{minipage}[t]{0.4\columnwidth}%
\textcolor{white}{.}\\
 \includegraphics[scale=0.7]{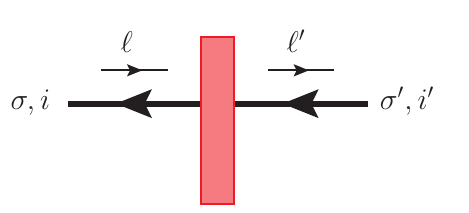}%
\end{minipage} & %
\begin{minipage}[t]{0.55\columnwidth}%
The interaction vertex of the antiquark with the shock wave: 
\[
T_{\sigma,\sigma';\,i,i'}^{\bar{Q}}=-2\pi\delta\left(\ell^{+}-\ell'{}^{+}\right)\gamma_{\sigma,\,\sigma'}^{+}\int d^{2}\boldsymbol{z}e^{i\left(\boldsymbol{\ell}-\boldsymbol{\ell}'\right)\cdot\boldsymbol{z}}\left[U^{\dagger}\left(\boldsymbol{z}\right)-1\right]_{i,i'}
\]
\end{minipage}\tabularnewline
\hline 
\begin{minipage}[t]{0.4\columnwidth}%
\textcolor{white}{.}\\
 \includegraphics[scale=0.7]{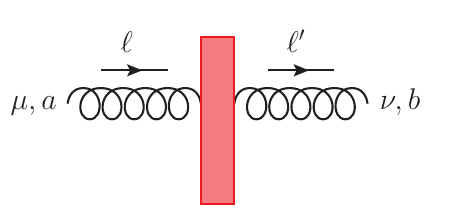}%
\end{minipage} & %
\begin{minipage}[t]{0.55\columnwidth}%
The interaction vertex of the gluon with the shock wave: 
\[
T_{\mu,\nu;a,b}^{g}=-2\pi\delta\left(\ell^{+}-\ell'{}^{+}\right)g_{\mu\nu}{\rm sgn}\left(\ell^{+}\right)\times
\]
\[
\quad\quad\times\int d^{2}\boldsymbol{z}e^{-i\left(\boldsymbol{\ell}'-\boldsymbol{\ell}\right)\cdot\boldsymbol{z}}\left[\mathcal{U}^{{\rm sgn\left(\ell^{+}\right)}}\left(\boldsymbol{z}\right)-1\right]_{b,a}
\]
\end{minipage}\tabularnewline
\hline 
\end{tabular}\caption{The CGC Feynman rules for the interaction of the partons with the
shock wave (vertical red-colored block)~~\cite{Blaizot:2004,Ayala:2017rmh,Caucal:2021ent,Caucal:2022ulg}.
We use the notations $\ell,\ell'$ for the momenta of the partons
before and after interaction with the shock wave, $\sigma,\sigma'$
for Dirac indices of the quarks, $i,\,i'$ for the color indices in
the fundamental representations $\boldsymbol{3}$ or $\bar{\boldsymbol{3}}$,
and $a,\,b$ for the gluon color indices in the adjoint irreducible
representation $\boldsymbol{8}$ of the color group. The matrices
$U$ and $\mathcal{U}$ are the Wilson lines in the fundamental and
the adjoint representations, respectively. }
\label{tab:FR}
\end{table}

In general the Wilson line~(\ref{eq:Wilson}) depends on color charge
configurations $\rho(x^{-},\boldsymbol{x}_{\perp})$, whose probability
in the CGC framework is described by the target-dependent weight functional
$W[\rho]$~\cite{McLerran:1993ni,McLerran:1993ka,McLerran:1994vd}.
The physical observables can be obtained averaging over all possible
configurations $\rho(x^{-},\boldsymbol{x}_{\perp})$, namely 
\begin{equation}
\left\langle \mathcal{A}\right\rangle =\int\mathcal{D}\rho\,W[\rho]\,\mathcal{A}[\rho],\label{eq:Ave}
\end{equation}
where $\mathcal{A[\rho]}$ corresponds to observable found for a fixed
distribution of charges $\rho(x^{-},\boldsymbol{x}_{\perp})$, and
the angular brackets $\langle...\rangle$ imply the above-mentioned
averaging. The analytic evaluation of the integral $\mathcal{D}\rho$
is possible only for simple forms of $W[\rho]$, like e.g. for the
gaussian parametrization implemented in~\cite{McLerran:1993ni,McLerran:1993ka}.
Fortunately, many observables can be expressed via universal target-dependent
correlators of several Wilson lines, which can be interpreted as multipole
scattering amplitudes and can be extracted them from phenomenological
analysis. For example, in many processes which involve production
of heavy quark and antiquark, the interaction with the target after
averaging over the color charges~(\ref{eq:Ave}) reduces to the nonperturbative
dipole $S$-matrix element~\cite{Gelis:2010nm,Kovchegov:2012mbw,Iancu:2003uh}

\begin{equation}
S_{2}\left(Y,\,\boldsymbol{x}_{q},\,\boldsymbol{x}_{\bar{q}}\right)=\frac{1}{N_{c}}\left\langle {\rm tr}\left(U\left(\boldsymbol{\boldsymbol{x}}_{q}\right)U^{\dagger}\left(\boldsymbol{x}_{\bar{q}}\right)\right)\right\rangle _{Y},\label{eq:S_matrix-2}
\end{equation}
where we use notation $Y$ for the rapidity of the dipole. The phenomenological
parametrizations of $S_{2}\left(Y,\,\boldsymbol{x}_{q},\,\boldsymbol{x}_{\bar{q}}\right)$
frequently are given in terms of the dipole scattering amplitude $N(x,\,\boldsymbol{r},\,\boldsymbol{b})$,
which is related to $S_{2}\left(Y,\,\boldsymbol{x}_{q},\,\boldsymbol{x}_{\bar{q}}\right)$
as 
\begin{equation}
N\left(x,\,\boldsymbol{r},\,\boldsymbol{b}\right)=1-S_{2}\left(Y=\ln\left(\frac{1}{x}\right),\,\boldsymbol{x}_{q},\,\boldsymbol{x}_{\bar{q}}\right),\label{eq:NS-1}
\end{equation}
where we introduced the variables $\boldsymbol{r}\equiv\boldsymbol{x}_{q}-\boldsymbol{x}_{\bar{q}}$
and $\boldsymbol{b}\equiv\alpha_{q}\,\boldsymbol{x}_{q}+\alpha_{\bar{q}}\boldsymbol{x}_{\bar{q}}$
for the transverse size and the transverse position of the dipole's
center of mass. Similarly, the processes which involve multiple heavy
quarks are sensitive to the multipole scattering $S$-matrix elements
(correlators of Wilson lines)
\begin{equation}
S_{2n}(Y,\,\boldsymbol{x}_{1},\,\boldsymbol{\xi}_{1},...,\boldsymbol{x}_{n},\,\boldsymbol{\xi}_{n})=\frac{1}{N_{c}}\left\langle {\rm tr}\left(U\left(\boldsymbol{\boldsymbol{x}}_{1}\right)U^{\dagger}\left(\boldsymbol{\xi}_{1}\right)...U\left(\boldsymbol{\boldsymbol{x}}_{n}\right)U^{\dagger}\left(\boldsymbol{\xi}_{n}\right)\right)\right\rangle _{Y},\label{eq:MultipoleDfinition}
\end{equation}
which correspond to the scattering amplitudes of $2n$-particle ensemble
of quarks. Such correlators represent important characteristics of
the target related to the gluon distributions of the latter. The higher
order correlators reduce to the correlators of inferior order in some
special limit (for example, when the neighbor arguments coincide,
$\boldsymbol{x}_{i}=\boldsymbol{\xi}_{i}$ or $\boldsymbol{x}_{i}=\boldsymbol{\xi}_{i-1}$).
However the inverse statement in general is not true, though may happen
in some models.

\begin{figure}
\includegraphics[width=9cm]{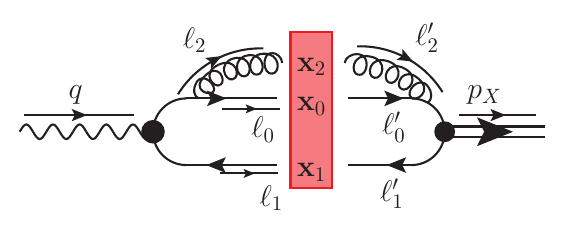}

\caption{\protect\label{fig:CGCBasic-2-1}A typical diagram which shows interaction
of the quark, antiquark and gluon with the shockwave (we use notations
with respective subindices $0,1,2$ to distinguish their coordinates).
The primed momenta ($\ell'_{0},\,\ell'_{1},\,\ell'_{2}$) correspond
to momenta after interaction with the shock wave (red block). The
dummy variables $\boldsymbol{x}_{0},\,\boldsymbol{x}_{1},\,\boldsymbol{x}_{2}$
are the configuration-space transverse coordinates of the partons
during interaction with the shock wave. See the text for a more detailed
explanation. }
\end{figure}

Finally, we would like to mention that the structure of the shockwave
vertices allows to represent the corresponding perturbative impact
factors, which appear in front of different correlators $S_{2n}(Y,\,\boldsymbol{x}_{1},\,\boldsymbol{\xi}_{1},...,\boldsymbol{x}_{n},\,\boldsymbol{\xi}_{n})$
in physical observables, as a convolution of the light cone wave functions
of the partonic ensemble of the initial and final states. Indeed,
let's consider for the sake of definiteness the diagram shown in the
Figure~\ref{fig:CGCBasic-2-1} (this process is not directly related
to heavy quarkonia pair production and is chosen merely for illustration
due to its simplicity; its extension to arbitrary number of partons
interacting with the shockwave is straightforward).

The matrices $U,U^{\dagger}$ are diagonal in helicity (Dirac) space,
and for this reason may be disregarded (factorized and taken out)
when analyzing Dirac algebra. The propagators of the free quarks and
antiquarks are given by 

\begin{equation}
S(k)=i\frac{\hat{k}+m}{k^{2}-m^{2}}.\label{eq:freeProp}
\end{equation}
and for onshell quarks ($k^{2}=m^{2}$) we may rewrite numerator using
conventional identities 
\begin{equation}
\sum_{h}u_{h}(k)\bar{u}_{h}(k)=\hat{k}+m,\qquad\sum_{h}v_{h}\left(\bar{k}\right)\bar{v}_{h}\left(\bar{k}\right)=\hat{k}-m
\end{equation}
For the propagators of the offshell fermions which pass through the
shockwave we may rewrite the numerators as 
\begin{align}
\hat{k}\mp m & =\left(\gamma^{+}\frac{\boldsymbol{k}_{\perp}^{2}+m^{2}}{2k^{+}}+\gamma^{-}k^{+}-\boldsymbol{\gamma}_{\perp}\cdot\boldsymbol{k}_{\perp}\mp m\right)+\gamma^{+}\left(k^{-}-\frac{\boldsymbol{k}_{\perp}^{2}+m^{2}}{2k^{+}}\right)=\\
 & =\left\{ \begin{array}{c}
\sum_{h}u_{h}(\tilde{k})\bar{u}_{h}(\tilde{k})\\
\sum_{\bar{h}}v_{\bar{h}}(\tilde{k})\bar{v}_{\bar{h}}(\tilde{k})
\end{array}\right.+\gamma^{+}\left(k^{-}-\frac{\boldsymbol{k}_{\perp}^{2}+m^{2}}{2k^{+}}\right),\qquad\qquad\qquad\tilde{k}=\left(k^{+},\frac{\boldsymbol{k}_{\perp}^{2}+m^{2}}{2k^{+}},\boldsymbol{k}_{\perp}\right).\nonumber 
\end{align}
Using the fact that the shockwave includes multiplication by $\gamma^{+}$
together with identity $\gamma^{+}\gamma^{+}=0$, we may see that
the last term $\sim\gamma^{+}$ in the last line does not contribute.
Finally, using the identities
\begin{equation}
\bar{u}_{h_{1}}\left(p_{1}\right)\gamma^{+}u_{h_{2}}\left(p_{2}\right)=\bar{v}_{h_{1}}\left(p_{1}\right)\gamma^{+}v_{h_{2}}\left(p_{2}\right)=2\sqrt{p_{1}^{+}p_{2}^{+}}\delta_{h_{1},h_{2}}
\end{equation}
we may obtain the set of identities
\begin{align}
\left(\hat{\ell}'+m\right) & T_{\,i,i'}^{Q}\left(\hat{\ell}+m\right)=\\
 & =2\pi\int d^{2}\boldsymbol{x}\left[U\left(\boldsymbol{x}\right)-1\right]_{i',i}\sum_{h}\,\,\left[u_{h}\left(\ell^{+},\frac{\left(\boldsymbol{\ell}'_{\perp}\right)^{2}+m^{2}}{2\ell^{+}},\,\boldsymbol{\ell}'_{\perp}\right)e^{-i\boldsymbol{\ell}'\cdot\boldsymbol{x}}\right]\,\left[\bar{u}_{h}\left(\ell^{+},\frac{\boldsymbol{\ell}_{\perp}^{2}+m^{2}}{2\ell^{+}},\,\boldsymbol{\ell}_{\perp}\right)e^{i\boldsymbol{\ell}\cdot\boldsymbol{x}}\right],\nonumber 
\end{align}
\begin{align}
\left(\hat{\ell'}-m\right) & T_{\,i,i'}^{\bar{Q}}\left(\hat{\ell}-m\right)=\\
 & =2\pi\int d^{2}\boldsymbol{x}\left[U^{\dagger}\left(\boldsymbol{x}\right)-1\right]_{i',i}\sum_{h}\,\,\left[v_{h}\left(\ell^{+},\frac{\left(\boldsymbol{\ell}'_{\perp}\right)^{2}+m^{2}}{2\ell^{+}},\,\boldsymbol{\ell}'_{\perp}\right)e^{-i\boldsymbol{\ell}'\cdot\boldsymbol{x}}\right]\,\left[\bar{v}_{h}\left(\ell^{+},\frac{\boldsymbol{\ell}_{\perp}^{2}+m^{2}}{2\ell^{+}},\,\boldsymbol{\ell}_{\perp}\right)e^{i\boldsymbol{\ell}\cdot\boldsymbol{x}}\right].\nonumber 
\end{align}
where the dummy integration variable $\boldsymbol{x}$ is the transverse
coordinate of the parton, and the arguments of the spinors $u_{h},v_{h},\bar{u}_{h},\bar{v}_{h}$
show explicitly the light-cone components of the corresponding partons.
This result can be used in order to demonstrate that in the mixed
light-cone representation in terms of the coordinates ($\ell^{+},\,\boldsymbol{x}$),
the interaction with the shockwave merely reduces to multiplication
by the factor $\left[U\left(\boldsymbol{x}\right)-1\right]$ for quarks
and $\left[U^{\dagger}\left(\boldsymbol{x}\right)-1\right]$ for antiquarks,
as expected in eikonal picture. It is possible to repeat this analysis
for the gluonic field, analyzing a convolution $\Pi_{\mu\alpha}^{ac}\left(\ell\right)T_{c,d}^{g,\alpha,\beta}\Pi_{\beta\nu}^{db}\left(\ell'\right)$,
where $\Pi_{\mu\nu}^{ab}$ is the light-cone propagator taken in the
Feynman or the light-cone gauge $A\cdot n=0$. It is straightforward
to show that in the light-cone representation the interaction of the
gluon with the shock wave reduces to multiplication by the factor
$\left[\mathcal{U}^{{\rm sgn\left(\ell^{+}\right)}}\left(\boldsymbol{x}\right)-1\right]_{b,a}$,
where $\boldsymbol{x}$ is the correspondent transverse coordinate
of the gluon.

\subsection{Charmonia-bottomonia production via Single Parton Scattering }

\label{subsec:SPS} The single gluon in the initial state may produce
a heavy charmonia bottomonia pairs via two complementary mechanisms,
a hadronization of $\bar{Q}Q$ pairs into final-state quarkonia (which
is described in NRQCD framework) and the fragmentation mechanism,
when each of the heavy quarkonia stems from fragmentation of only
one heavy (bottom) quark. The two mechanisms have completely different
final states and different underlying dynamics, and for this reason
will be disregarded in what follows.

\subsubsection{Production of quarkonia via $\bar{Q}Q$ fusion}

\begin{figure}
\includegraphics[width=4cm]{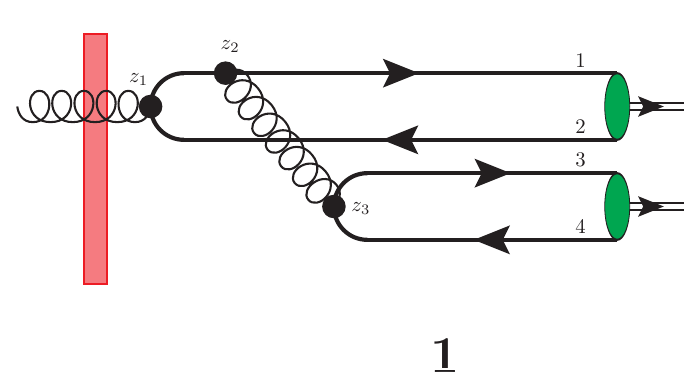}\includegraphics[width=4cm]{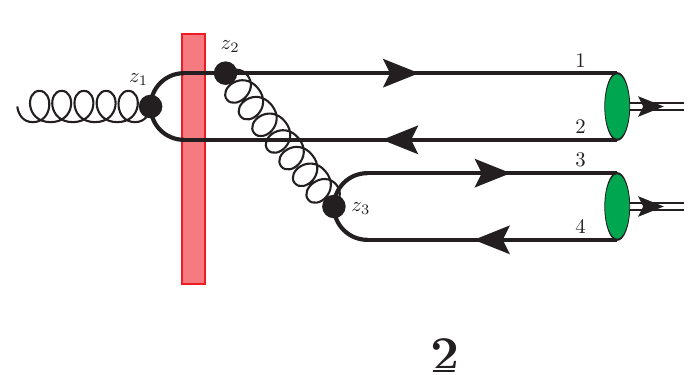}\includegraphics[width=4cm]{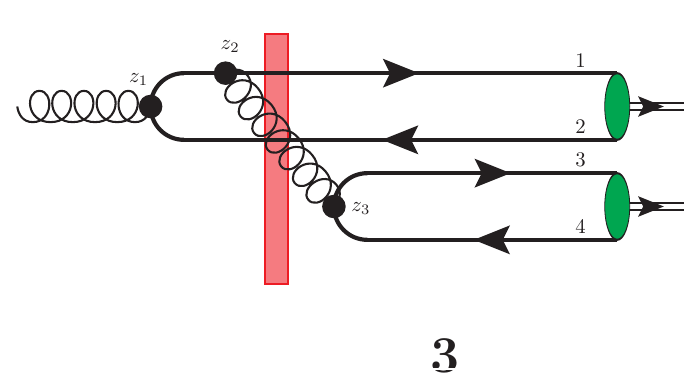}\includegraphics[width=4cm]{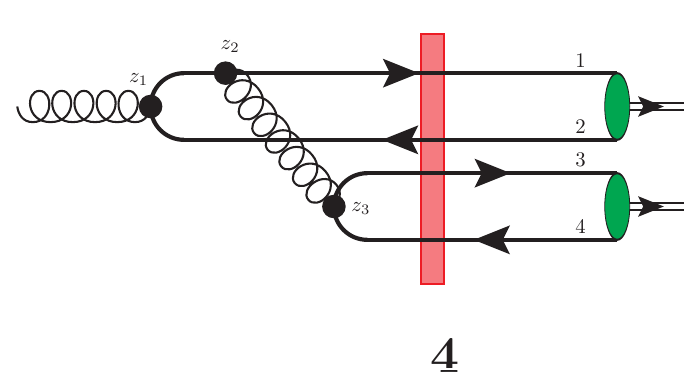}

\includegraphics[width=4cm]{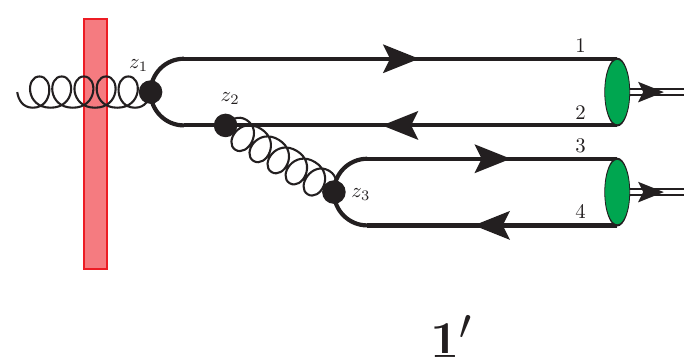}\includegraphics[width=4cm]{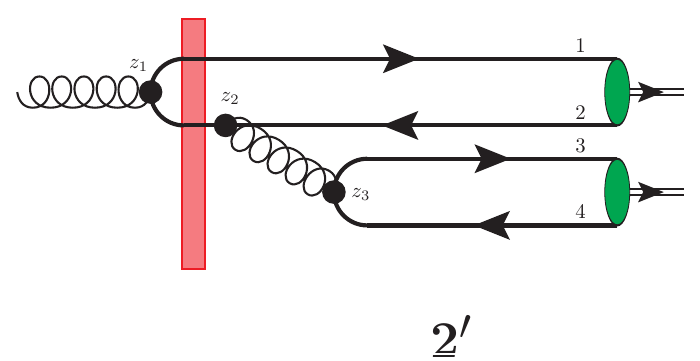}\includegraphics[width=4cm]{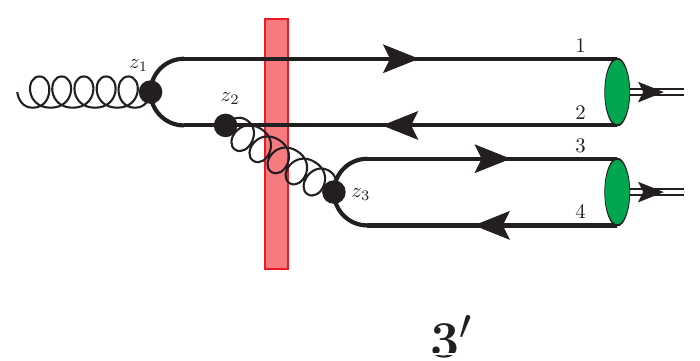}\includegraphics[width=4cm]{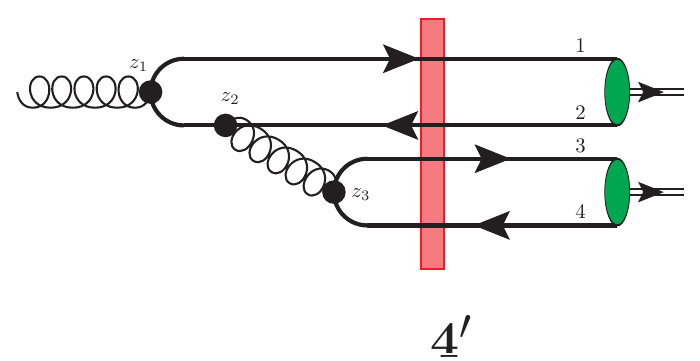}

\includegraphics[width=4cm]{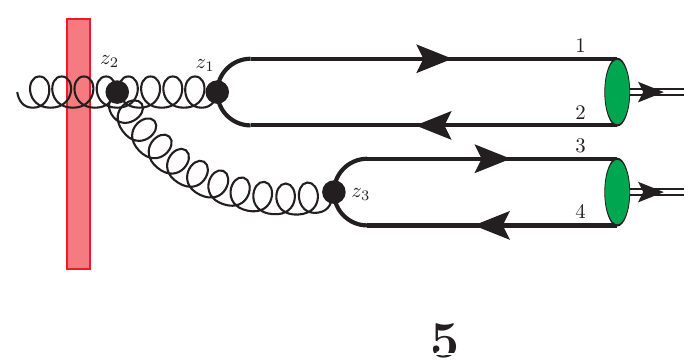}\includegraphics[width=4cm]{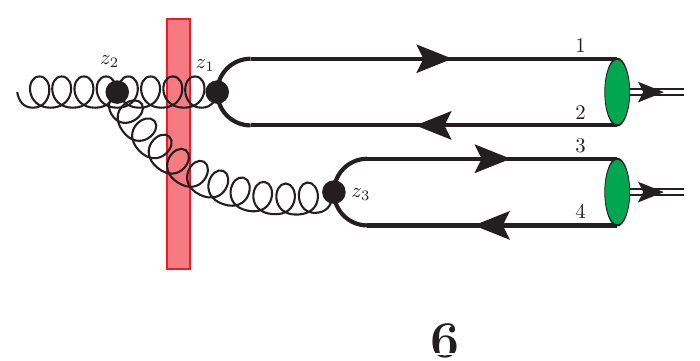}\includegraphics[width=4cm]{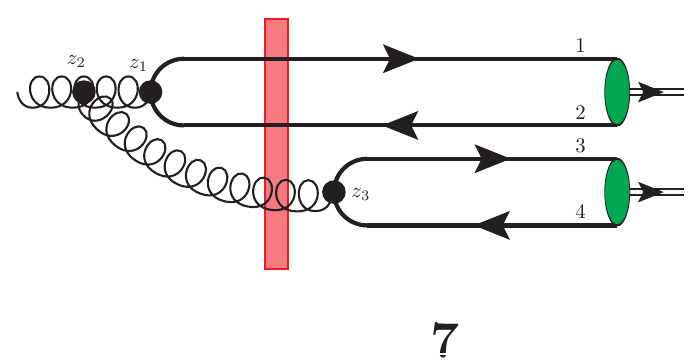}\includegraphics[width=4cm]{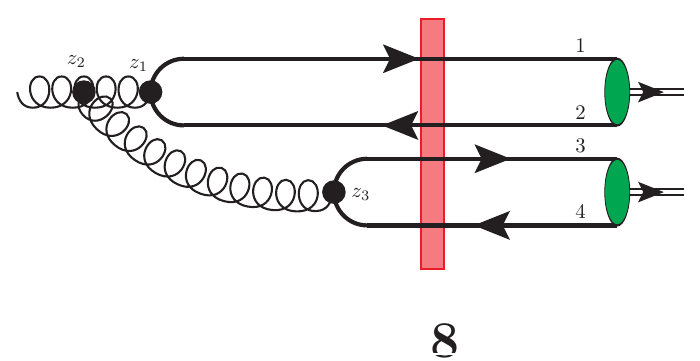}

\caption{\protect\label{fig:QuarkoniaPairCGC}The diagrams which describe
the inclusive charmonia-bottomonia pair production in CGC picture
in the leading order over $\alpha_{s}$. The diagrams 1'-4' in the
middle row are charge conjugate of the diagrams in the upper row (differ
by inversion of the upper quark loop). The diagrams in the first 3
columns require formation of the (lower) quarkonium from the color
octet $\bar{Q}Q$ (formed via $g\to\bar{Q}Q$ subprocess), and for
this reason are suppressed by a small color octet LDME. The dominant
color singlet contribution comes from the diagrams in the last row
(diagrams \textquotedblleft 4\textquotedblright , \textquotedblleft 4'~\textquotedblright{}
and \textquotedblleft 8\textquotedblright{} respectively). The subscript
numbers 1-4 in the right part of each diagram enumerate the heavy
quark lines in our convention. The subscript letters $z_{1}\,...\,z_{3}$
stand for the coordinates of the interaction vertices in the configuration
space. The red block represents the interaction with the target (shockwave).}
\end{figure}

\begin{figure}
\includegraphics[width=4cm]{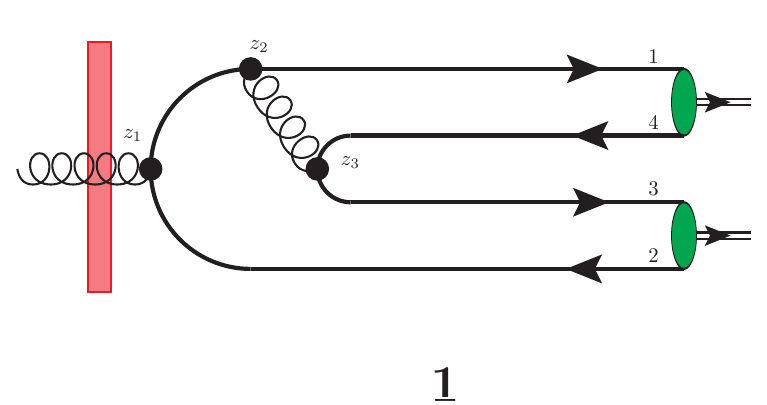}\includegraphics[width=4cm]{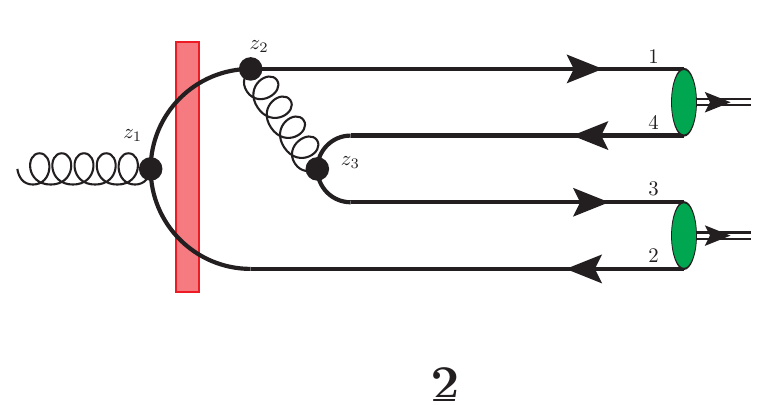}\includegraphics[width=4cm]{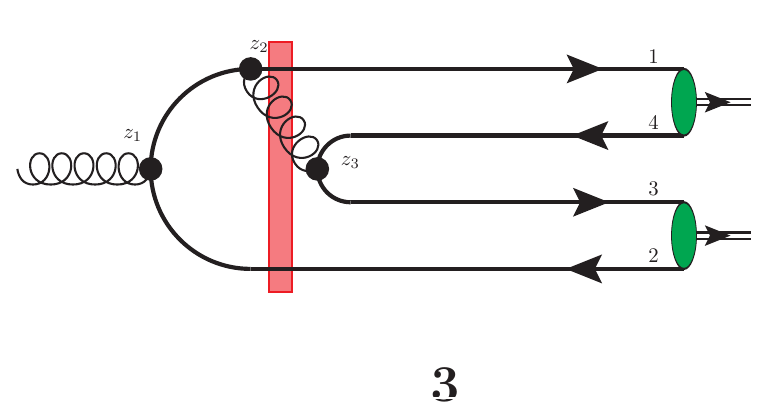}\includegraphics[width=4cm]{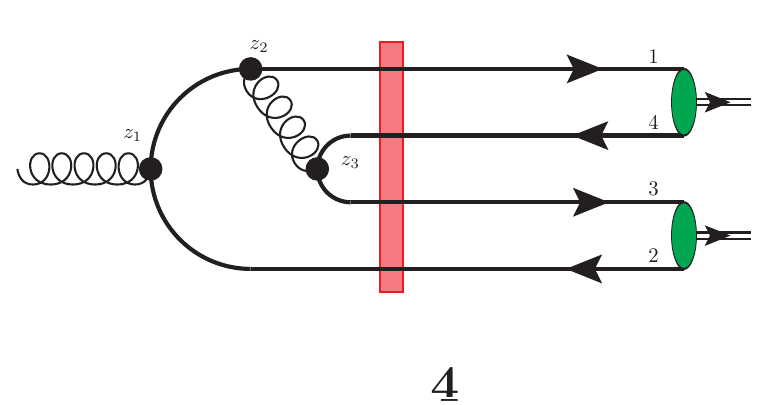}

\includegraphics[width=4cm]{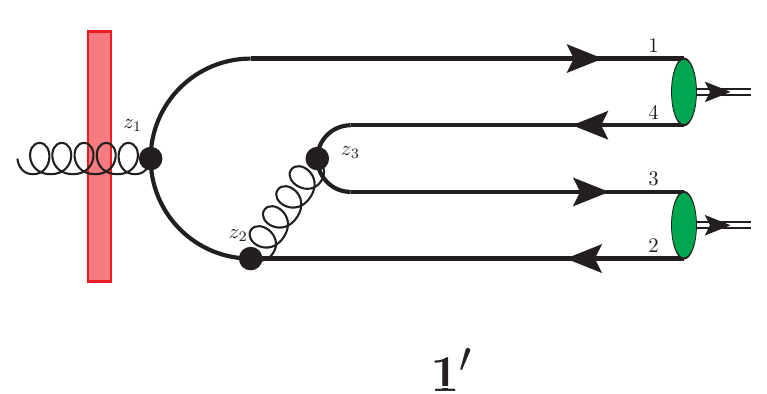}\includegraphics[width=4cm]{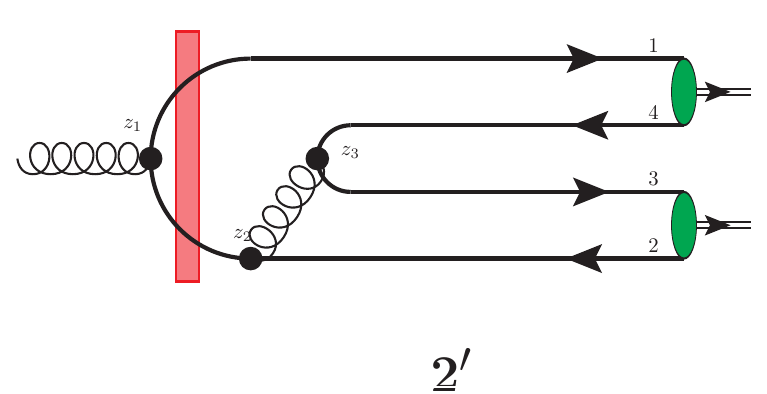}\includegraphics[width=4cm]{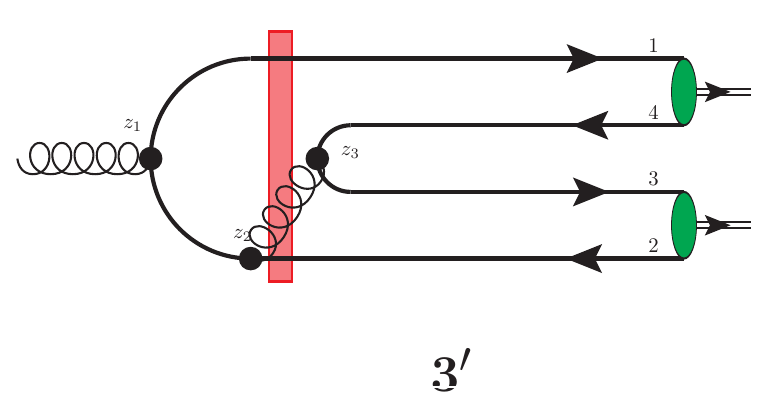}\includegraphics[width=4cm]{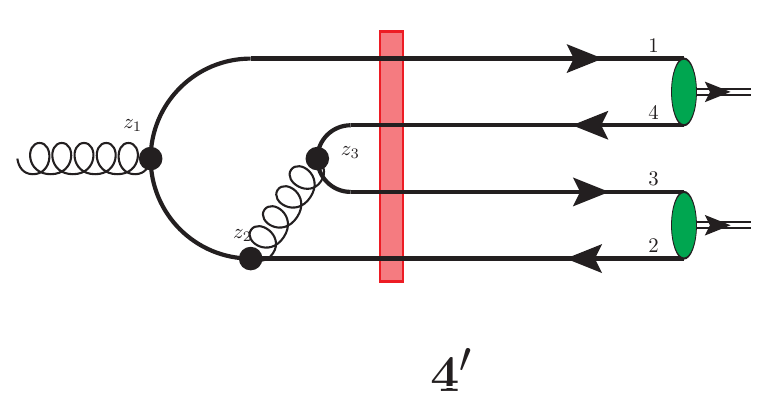}

\includegraphics[width=4cm]{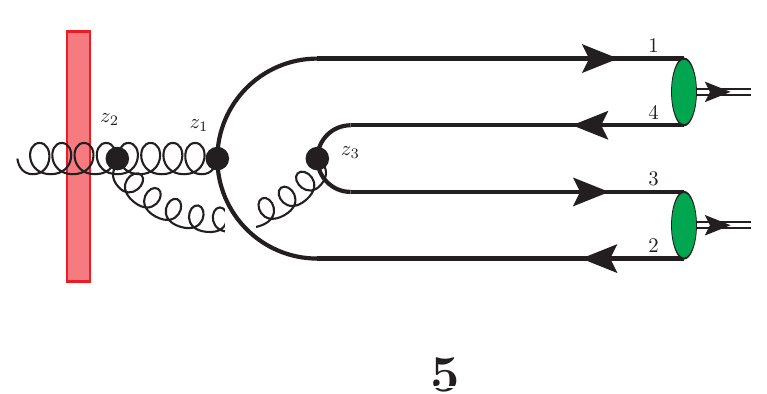}\includegraphics[width=4cm]{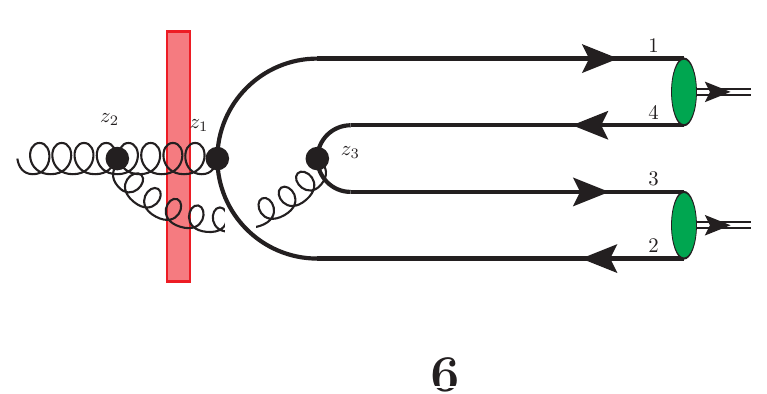}\includegraphics[width=4cm]{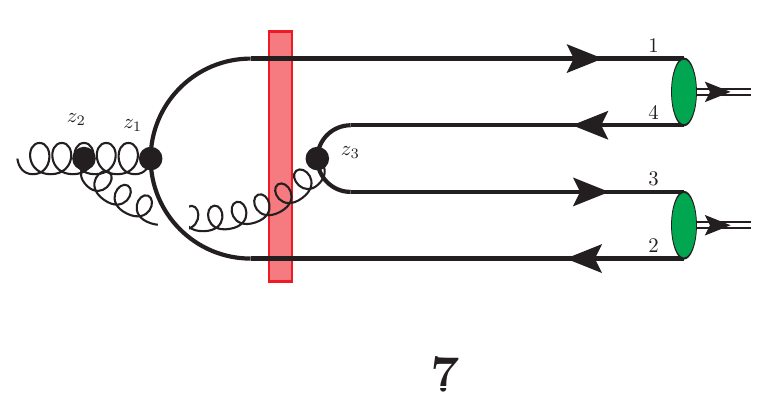}\includegraphics[width=4cm]{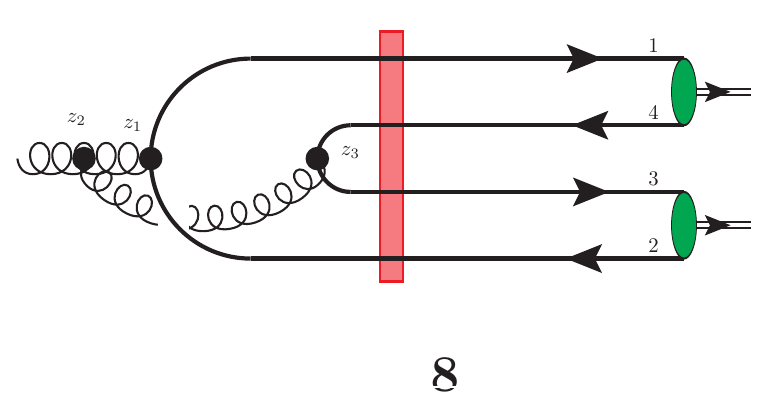}

\caption{\protect\label{fig:QuarkoniaPairCGC-1}Additional diagrams which in
general should be taken into account for inclusive quarkonia pair
production in CGC picture in the leading order over $\alpha_{s}$,
yet do not contribute to charmonia-bottomonia pairs. The diagrams
1-4 in the upper row and middle row are related by charge conjugation
(inverted direction of the quark lines). The diagrams in the first
column can proceed only via color octet LDMEs and thus are suppressed.
The subscript numbers 1-4 enumerate the heavy quark lines in our convention.
The subscript letters $z_{1}...z_{3}$ stand for the coordinates of
the interaction vertices in the configuration space. The red block
represents the interaction with the target (shockwave).}
\end{figure}

The charmonia-bottomonia pair production via single-parton scattering
proceeds via emission of the single gluon from the projectile and
subsequent formation of the heavy quarks in the field of the shock
wave (gluonic field of the target). The corresponding cross-section
is given by 

\begin{align}
\frac{d\sigma\left(p+p\to M_{1}+M_{2}+X\right)}{dy_{1}d^{2}\boldsymbol{p}_{1}^{\perp}dy_{2}d^{2}\boldsymbol{p}_{2}^{\perp}} & =x_{1}g\left(x_{1}\right)\frac{d\sigma\left(g+p\to M_{1}+M_{2}+X\right)}{dy_{1}d^{2}\boldsymbol{p}_{1}^{\perp}dy_{2}d^{2}\boldsymbol{p}_{2}^{\perp}}\label{eq:sixfold-1}
\end{align}
where $x_{1}g\left(x_{1}\right)$ is the standard forward gluon PDF
of the projectile, and $x_{1}=q^{+}/p_{{\rm projectile}}^{+}$is the
fraction of the projectile's light-cone ``+'' momentum carried by
the gluon. In what follows we will focus on the evaluation of the
second term in~(\ref{eq:sixfold-1}), which in the leading order
in strong coupling $\alpha_{s}$ obtains contributions from a set
of diagrams shown in the Figure~\ref{fig:QuarkoniaPairCGC}. We may
observe that the diagrams shown in the first three columns lead to
formation of one of the $\bar{Q}Q$ pairs (in inferior part of the
diagram) in the color octet state. Formally, the $S$-wave color octet
LDMEs are suppressed compared to color singlet LDMEs as $\sim\mathcal{O}\left(v^{4}\right)$,
where $v\sim\mathcal{O}\left(\alpha_{s}(m_{Q})\right)$ is the relative
velocity of the internal motion of heavy quarks inside the quarkonium,
and numerically the color octet LDME indeed are very small~\cite{Baranov:2016clx,Baranov:2019lhm}.
For this reason, in what follows we may safely disregard these contributions
and focus on evaluation of the diagrams in the last column. Furthermore,
since the shock wave implies exchange of the vacuum quantum numbers
in $t$-channel, and gluon has spin-1, the diagrams in the lower row
can contribute only if both produced quarkonia have vector quantum
numbers ($J^{P}=1^{-}$). For comparison, the production of quarkonia
pairs in general case requires inclusion of additional Feynman diagrams
shown in the Figure~\ref{fig:QuarkoniaPairCGC-1}, or emission of
additional hard gluons, leading to significantly more complicated
expressions for the cross-sections.

In view of the expected smallness of the cross-section, in what follows
we will focus on the kinematics of small transverse momenta $p_{T}$,
where most of the quarkonia pairs are produced. In this kinematics
we may safely disregard the diagrams in which the quarkonia formation
occurs from color octet quark-antiquark pair and thus is controlled
by small color octet LDMEs (diagrams 1-3, 1'-3', 5-7 in the Figure~\ref{fig:QuarkoniaPairCGC}).
For the remaining diagrams 4,~4',~8 the corresponding contribution
in eikonal picture can be described by a convolution of the so-called
perturbative impact factors with target-dependent correlators of multipole
matrix elements. The formation of the final-state quarkonia with spin
$S$ and spin projection $S_{z}$ from the $\bar{Q}Q$ is described
by the conventional NRQCD projectors~\cite{Bodwin:1994jh,Maltoni:1997pt,Brambilla:2008zg,Feng:2015cba,Brambilla:2010cs}
\begin{align}
\Pi_{M}^{SS_{Z}}\left(p_{M},\,k\right) & =\sqrt{\frac{\langle\mathcal{O}_{M}\rangle}{m}}\sum_{s,\bar{s}}\left\langle \frac{1}{2}s,\,\frac{1}{2}\bar{s}\bigg|S,S_{z}\right\rangle v_{\bar{s}}\left(p_{M}-\frac{k}{2}\right)\bar{u}_{s}\left(p_{M}+\frac{k}{2}\right)=\label{eq:NRQCD_Projector}\\
 & =\sqrt{\frac{\langle\mathcal{O}_{M}\rangle}{8m^{3}}}\left\{ \begin{array}{cc}
\left(\frac{\hat{p}_{M}}{2}-\hat{k}-m\right)\gamma_{5}\left(\frac{\hat{p}_{M}}{2}+\hat{k}+m\right), & \qquad S=0\\
\left(\frac{\hat{p}_{M}}{2}-\hat{k}-m\right)\hat{\varepsilon}^{*}\left(p_{M},\,S_{Z}\right)\left(\frac{\hat{p}_{M}}{2}+\hat{k}+m\right), & \qquad S=1
\end{array}\right.\nonumber 
\end{align}
where $\langle\mathcal{O}_{M}\rangle$ are the corresponding LDMEs,
and $\varepsilon^{\mu}\left(p_{M}\right)$ is the polarization vector
of the produced vector quarkonium which satisfies orthogonality condition
$\varepsilon\cdot p_{M}=0$. From now on we will use the notation
$m$ for the quark mass, and, in order to avoid confusion, will tacitly
replace it with $m_{1}$ and $m_{2}$ in expressions where both quarks
contribute. The momentum $k$ which describes the internal motion
of the quarks in the quarkonium can be disregarded in the heavy quark
mass limit, tacitly assuming that $k=0$ everywhere in what follows.
The evaluation of the remaining diagrams is straightforward in the
mixed (light-cone) representation introduced earlier. Using conventional
CGC rules for the vertices, we may obtain for the amplitudes of processes
shown in the diagrams 4,4',8

\begin{align}
\mathcal{A}_{4}^{(a)} & =\int\prod_{n=1}^{4}\left(d^{2}\boldsymbol{x}_{n}\right)e^{-i\boldsymbol{p}_{1}\cdot\left(\boldsymbol{x}_{1}+\boldsymbol{x}_{2}\right)/2}e^{-i\boldsymbol{p}_{2}\cdot\left(\boldsymbol{x}_{3}+\boldsymbol{x}_{4}\right)/2}\,R_{4}\left(\boldsymbol{x}_{1},\,\boldsymbol{x}_{2},\,\boldsymbol{x}_{3},\,\boldsymbol{x}_{4}\right)T_{a}^{(4)}\left(\boldsymbol{x}_{1},\,\boldsymbol{x}_{2},\,\boldsymbol{x}_{3},\,\boldsymbol{x}_{4}\right)\\
\mathcal{A}_{4'}^{(a)} & =\int\prod_{n=1}^{4}\left(d^{2}\boldsymbol{x}_{n}\right)e^{-i\boldsymbol{p}_{1}\cdot\left(\boldsymbol{x}_{1}+\boldsymbol{x}_{2}\right)/2}e^{-i\boldsymbol{p}_{2}\cdot\left(\boldsymbol{x}_{3}+\boldsymbol{x}_{4}\right)/2}\,R_{4'}\left(\boldsymbol{x}_{1},\,\boldsymbol{x}_{2},\,\boldsymbol{x}_{3},\,\boldsymbol{x}_{4}\right)T_{a}^{(4')}\left(\boldsymbol{x}_{1},\,\boldsymbol{x}_{2},\,\boldsymbol{x}_{3},\,\boldsymbol{x}_{4}\right)\\
\mathcal{A}_{8}^{(a)} & =\int\prod_{n=1}^{4}\left(d^{2}\boldsymbol{x}_{n}\right)e^{-i\boldsymbol{p}_{1}\cdot\left(\boldsymbol{x}_{1}+\boldsymbol{x}_{2}\right)/2}e^{-i\boldsymbol{p}_{2}\cdot\left(\boldsymbol{x}_{3}+\boldsymbol{x}_{4}\right)/2}\,R_{8}\left(\boldsymbol{x}_{1},\,\boldsymbol{x}_{2},\,\boldsymbol{x}_{3},\,\boldsymbol{x}_{4}\right)T_{a}^{(8)}\left(\boldsymbol{x}_{1},\,\boldsymbol{x}_{2},\,\boldsymbol{x}_{3},\,\boldsymbol{x}_{4}\right)
\end{align}
where $a$ is the color index of the incoming gluon, $R_{4},\,R_{4}',\,R_{8}$
are the perturbative impact factors, and $T_{a}^{(4)},\,T_{a}^{(4')},\,T_{a}^{(8)}$
are the target-dependent (multipole) correlators of Wilson lines.
Explicitly, the impact factors are given by 
\begin{align}
R_{4}\left(\boldsymbol{x}_{1},\,\boldsymbol{x}_{2},\,\boldsymbol{x}_{3},\,\boldsymbol{x}_{4}\right) & =\left(ig\right)^{3}\int\prod_{n=1}^{4}\left(\frac{d^{3}\ell_{n}}{\left(2\pi\right)^{3}}e^{i\boldsymbol{\ell}_{n}\cdot\boldsymbol{x}_{n}}\right){\rm tr}\left[\Pi_{M_{1}}^{SS_{Z}}\left(p_{1},\,0\right)\gamma^{+}S\left(\ell_{1}\right)\gamma^{\mu}S\left(\ell_{1}-\ell_{3}-\ell_{4}\right)\hat{\varepsilon}\left(q\right)S\left(-\ell_{2}\right)\gamma_{+}\right]\times\label{eq:R4}\\
 & \times{\rm \Pi_{\mu\nu}\left(\ell_{3}+\ell_{4}\right)tr}\left[\Pi_{M_{2}}^{SS_{Z}}\left(p_{2},0\right)\gamma^{+}S\left(\ell_{3}\right)\gamma^{\nu}S\left(-\ell_{4}\right)\gamma_{+}\right]\left(2\pi\right)^{3}\delta^{3}\left(q-\sum_{n}\ell_{n}\right)\nonumber 
\end{align}
\begin{align}
R_{4'}\left(\boldsymbol{x}_{1},\,\boldsymbol{x}_{2},\,\boldsymbol{x}_{3},\,\boldsymbol{x}_{4}\right) & =\left(ig\right)^{3}\int\prod_{n=1}^{4}\left(\frac{d^{3}\ell_{n}}{\left(2\pi\right)^{3}}e^{i\boldsymbol{\ell}_{n}\cdot\boldsymbol{x}_{n}}\right){\rm tr}\left[\Pi_{M_{1}}^{SS_{Z}}\left(p_{1},\,0\right)\gamma^{+}S\left(\ell_{1}\right)\hat{\varepsilon}\left(q\right)S\left(\ell_{1}-q\right)\gamma^{\mu}S\left(-\ell_{2}\right)\gamma_{+}\right]\times\\
 & \times{\rm \Pi_{\mu\nu}\left(\ell_{3}+\ell_{4}\right)tr}\left[\Pi_{M_{2}}^{SS_{Z}}\left(p_{2},0\right)\gamma^{+}S\left(\ell_{3}\right)\gamma^{\nu}S\left(-\ell_{4}\right)\gamma_{+}\right]\left(2\pi\right)^{3}\delta^{3}\left(q-\sum_{n}\ell_{n}\right)\nonumber 
\end{align}

\begin{align}
R_{8}\left(\boldsymbol{x}_{1},\,\boldsymbol{x}_{2},\,\boldsymbol{x}_{3},\,\boldsymbol{x}_{4}\right) & =\left(ig\right)^{3}\int\prod_{n=1}^{4}\left(\frac{d^{3}\ell_{n}}{\left(2\pi\right)^{3}}e^{i\boldsymbol{\ell}_{n}\cdot\boldsymbol{x}_{n}}\right)\varepsilon^{\lambda}(q)V_{\lambda\alpha\beta}\Pi_{\alpha\mu}\left(\ell_{1}+\ell_{2}\right){\rm tr}\left[\Pi_{M_{1}}^{SS_{Z}}\left(p_{1},\,0\right)\gamma^{+}S\left(\ell_{1}\right)\gamma^{\mu}S\left(-\ell_{2}\right)\gamma_{+}\right]\times\label{eq:R8}\\
 & \times{\rm \Pi_{\beta\nu}\left(\ell_{3}+\ell_{4}\right)tr}\left[\Pi_{M_{2}}^{SS_{Z}}\left(p_{2},0\right)\gamma^{+}S\left(\ell_{3}\right)\gamma^{\nu}S\left(-\ell_{4}\right)\gamma_{+}\right]\left(2\pi\right)^{3}\delta^{3}\left(q-\sum_{n}\ell_{n}\right)\nonumber 
\end{align}
where the notation ${\rm tr(...)}$ implies that the trace should
be taken only over the Dirac indices, $\Pi_{\mu\nu}(\ell)$ is the
color-independent part of the gluon propagator, and the momenta $\ell_{a}$
are the momenta of the quarks before interaction with the shockwave.
According to CGC rules discussed in Section~\ref{subsec:Derivation}
the interaction with the shock wave does not change the plus-component
of the quark momentum, and for this reason the plus-components $\ell_{n}^{+}$
of the momenta are fully determined by the kinematics of the corresponding
meson: they coincide with $p_{1}^{+}/2$ for $n=1,2$ and with $p_{2}^{+}/2$
for $n=3,4$. For this reason, integrals over the quark momenta include
only 3 components, namely $d^{3}\ell_{n}=d\ell_{n}^{-}d^{2}\boldsymbol{\ell}_{n}$.
The $\delta^{3}(...)$ functions which appear in the integrands of~(\ref{eq:R4}-\ref{eq:R8})
reflect conservation of the total momentum of the partonic ensemble.
The notation $V_{\lambda\alpha\beta}$ in~(\ref{eq:R8}) is the colorless
part of the standard perturbative 3-gluon vertex,
\begin{equation}
V_{\lambda\alpha\beta}=g^{\lambda\alpha}\left(q-\ell_{1}-\ell_{2}\right)^{\beta}-g^{\lambda\beta}\left(q-\ell_{3}-\ell_{4}\right)^{\beta}+g^{\alpha\beta}\left(\ell_{1}+\ell_{2}-\ell_{3}-\ell_{4}\right)^{\lambda}.
\end{equation}
The target-dependent factors $T_{a}^{(4)},\,T_{a}^{(4')},\,T_{a}^{(8)}$
are given explicitly by
\begin{align}
T_{a}^{(4)}\left(\boldsymbol{x}_{1},\,\boldsymbol{x}_{2},\,\boldsymbol{x}_{3},\,\boldsymbol{x}_{4}\right) & ={\rm tr}_{c}\left(t_{a}t_{b}U\left(\boldsymbol{x}_{1}\right)U^{\dagger}\left(\boldsymbol{x}_{2}\right)\right){\rm tr}_{c}\left(t_{b}U\left(\boldsymbol{x}_{3}\right)U^{\dagger}\left(\boldsymbol{x}_{4}\right)\right)=\label{eq:T4}\\
 & =\frac{N_{c}}{2}\left\langle S_{4}^{(a)}\left(\boldsymbol{x}_{3},\,\boldsymbol{x}_{4},\,\boldsymbol{x}_{1},\,\boldsymbol{x}_{2}\right)-S_{2}^{(a)}\left(\boldsymbol{x}_{1},\,\boldsymbol{x}_{2}\right)S_{2}\left(\boldsymbol{x}_{3},\,\boldsymbol{x}_{4}\right)\right\rangle \nonumber 
\end{align}
\begin{align}
T_{a}^{(4')}\left(\boldsymbol{x}_{1},\,\boldsymbol{x}_{2},\,\boldsymbol{x}_{3},\,\boldsymbol{x}_{4}\right) & ={\rm tr}_{c}\left(t_{b}t_{a}U\left(\boldsymbol{x}_{1}\right)U^{\dagger}\left(\boldsymbol{x}_{2}\right)\right){\rm tr}_{c}\left(t_{b}U\left(\boldsymbol{x}_{3}\right)U^{\dagger}\left(\boldsymbol{x}_{4}\right)\right)=\label{eq:T4Prime}\\
 & =\frac{N_{c}}{2}\left\langle S_{4}^{(a)}\left(\boldsymbol{x}_{1},\,\boldsymbol{x}_{2},\,\boldsymbol{x}_{3},\,\boldsymbol{x}_{4}\right)-S_{2}^{(a)}\left(\boldsymbol{x}_{1},\,\boldsymbol{x}_{2}\right)S_{2}\left(\boldsymbol{x}_{3},\,\boldsymbol{x}_{4}\right)\right\rangle \nonumber 
\end{align}
\begin{align}
T_{a}^{(8)}\left(\boldsymbol{x}_{1},\,\boldsymbol{x}_{2},\,\boldsymbol{x}_{3},\,\boldsymbol{x}_{4}\right) & =if_{abc}{\rm tr}_{c}\left(t_{b}U\left(\boldsymbol{x}_{1}\right)U^{\dagger}\left(\boldsymbol{x}_{2}\right)\right){\rm tr}_{c}\left(t_{c}U\left(\boldsymbol{x}_{3}\right)U^{\dagger}\left(\boldsymbol{x}_{4}\right)\right)=\label{eq:T8}\\
 & =\frac{N_{c}}{2}\left\langle S_{4}^{(a)}\left(\boldsymbol{x}_{3},\,\boldsymbol{x}_{4},\,\boldsymbol{x}_{1},\,\boldsymbol{x}_{2}\right)-S_{4}^{(a)}\left(\boldsymbol{x}_{1},\,\boldsymbol{x}_{2},\,\boldsymbol{x}_{3},\,\boldsymbol{x}_{4}\right)\right\rangle \nonumber \\
 & =T_{a}^{(4')}\left(\boldsymbol{x}_{1},\,\boldsymbol{x}_{2},\,\boldsymbol{x}_{3},\,\boldsymbol{x}_{4}\right)-T_{a}^{(4)}\left(\boldsymbol{x}_{1},\,\boldsymbol{x}_{2},\,\boldsymbol{x}_{3},\,\boldsymbol{x}_{4}\right)\nonumber 
\end{align}
where the notation ${\rm tr_{c}(...)}$ implies that the trace should
be taken only over the color indices, and we introduced shorthand
notations 
\begin{align}
S_{4}^{(a)}\left(\boldsymbol{x}_{1},\,\boldsymbol{x}_{2},\,\boldsymbol{x}_{3},\,\boldsymbol{x}_{4}\right)= & \frac{1}{N_{c}}\left\langle {\rm tr}_{c}\left({\color{red}t_{a}}U\left(\boldsymbol{\boldsymbol{x}}_{1}\right)U^{\dagger}\left(\boldsymbol{\boldsymbol{x}}_{2}\right)U\left(\boldsymbol{\boldsymbol{x}}_{3}\right)U^{\dagger}\left(\boldsymbol{\boldsymbol{x}}_{4}\right)\right)\right\rangle _{Y},\\
S_{2}^{(a)}\left(\boldsymbol{x}_{1},\,\boldsymbol{x}_{2}\right)= & \frac{1}{N_{c}}\left\langle {\rm tr}_{c}\left({\color{red}t_{a}}U\left(\boldsymbol{\boldsymbol{x}}_{1}\right)U^{\dagger}\left(\boldsymbol{\boldsymbol{x}}_{2}\right)\right)\right\rangle _{Y}.
\end{align}
For derivation of~(\ref{eq:T4}-\ref{eq:T8}) we used heavily the
Fierz identities for Gell-Mann matrices 
\begin{equation}
\left(t_{b}\right)_{c_{1}}^{c_{1}'}\left(t_{b}\right)_{c_{2}}^{c_{2}'}=\delta_{c_{2}}^{c_{1}'}\delta_{c_{1}}^{c_{2}'}-\frac{1}{N_{c}}\delta_{c_{1}}^{c_{1}'}\delta_{c_{2}}^{c_{2}'},\qquad t_{b}{\rm tr}_{c}\left(t_{b}\hat{A}\right)=\frac{1}{2}\left[\hat{A}-\frac{\hat{\boldsymbol{1}}}{N_{c}}{\rm tr}\left(A\right)\right],\,\,\forall A,\label{eq:Fierz}
\end{equation}
and rewrote $if_{abc}$ as 
\begin{equation}
if_{abc}\equiv2i{\rm tr}\left[t_{a}\left(t_{b}t_{c}-t_{c}t_{b}\right)\right]
\end{equation}
 in order to obtain the results shown in the second line of~(\ref{eq:T8}).
The relation which appears in the last line of~(\ref{eq:T8}) implies
that the expression should be antisymmetric w.r.t. permutation of
mesons, and indicates that it will give a numerically small contribution.

The evaluation of the impact factors $R_{4},R_{4'},R_{8}$ simplifies
drastically using the procedure discussed at the end of the Section~\ref{subsec:Derivation}.
Replacing the propagators connected to the shockwave with 
\begin{equation}
\frac{i\sum_{h}u_{h}(\tilde{\ell})\bar{u}_{h}(\tilde{\ell})}{\ell^{2}-m^{2}+i0},\qquad{\rm and}\qquad\frac{i\sum_{h}v_{h}(\tilde{\ell})\bar{v}_{h}(\tilde{\ell})}{\ell^{2}-m^{2}+i0},\qquad\tilde{\ell}=\left(\left|\ell^{+}\right|,\frac{\boldsymbol{\ell}_{\perp}^{2}+m^{2}}{2\left|\ell^{+}\right|},\boldsymbol{\ell}_{\perp}\right)\label{eq:onshell}
\end{equation}
for quarks and antiquarks respectively, we may obtain for $R_{4},\,R_{4'}$
\begin{align}
R_{4}\left(\boldsymbol{x}_{1},\,\boldsymbol{x}_{2},\,\boldsymbol{x}_{3},\,\boldsymbol{x}_{4}\right) & =\left(ig\right)^{3}\sum_{h_{1},...,h_{4}}\int\prod_{n=1}^{4}\left(\frac{d^{3}\ell_{n}}{\left(2\pi\right)^{3}}e^{i\boldsymbol{\ell}_{n}\cdot\boldsymbol{x}_{n}}\right)\Pi_{M_{1}}^{h_{1},h_{2}}\Pi_{M_{2}}^{h_{3},h_{4}}\left(2\pi\right)^{3}\delta^{3}\left(q-\sum_{n}\ell_{n}\right)\times\label{eq:R4-1}\\
 & \times\left[\frac{\bar{u}_{h_{1}}\left(\tilde{\ell}_{1}\right)\gamma^{\mu}S\left(\ell_{1}-\ell_{3}-\ell_{4}\right)\hat{\varepsilon}\left(q\right)v_{h_{2}}\left(\tilde{\ell}_{2}\right)}{\left(\ell_{1}^{2}-m_{1}^{2}+i0\right)\left(\ell_{2}^{2}-m_{1}^{2}+i0\right)}\right]{\rm \Pi_{\mu\nu}\left(\ell_{3}+\ell_{4}\right)}\left[\frac{\bar{u}_{h_{3}}\left(\tilde{\ell}_{3}\right)\gamma^{\nu}v_{h_{4}}\left(\tilde{\ell}_{4}\right)}{\left(\ell_{3}^{2}-m_{2}^{2}+i0\right)\left(\ell_{4}^{2}-m_{2}^{2}+i0\right)}\right]\nonumber 
\end{align}
\begin{align}
R_{4'}\left(\boldsymbol{x}_{1},\,\boldsymbol{x}_{2},\,\boldsymbol{x}_{3},\,\boldsymbol{x}_{4}\right) & =\left(ig\right)^{3}\sum_{h_{1},...,h_{4}}\int\prod_{n=1}^{4}\left(\frac{d^{3}\ell_{n}}{\left(2\pi\right)^{3}}e^{i\boldsymbol{\ell}_{n}\cdot\boldsymbol{x}_{n}}\right)\Pi_{M_{1}}^{h_{1},h_{2}}\Pi_{M_{2}}^{h_{3},h_{4}}\left(2\pi\right)^{3}\delta^{3}\left(q-\sum_{n}\ell_{n}\right)\times\label{eq:R4Prime-1}\\
 & \times\left[\frac{\bar{u}_{h_{1}}\left(\tilde{\ell}_{1}\right)\hat{\varepsilon}\left(q\right)S\left(\ell_{1}-q\right)\gamma^{\mu}v_{h_{2}}\left(\tilde{\ell}_{2}\right)}{\left(\ell_{1}^{2}-m_{1}^{2}+i0\right)\left(\ell_{2}^{2}-m_{1}^{2}+i0\right)}\right]\Pi_{\mu\nu}\left(\ell_{3}+\ell_{4}\right)\left[\frac{\bar{u}_{h_{3}}\left(\tilde{\ell}_{3}\right)\gamma^{\nu}v_{h_{4}}\left(\tilde{\ell}_{4}\right)}{\left(\ell_{3}^{2}-m_{2}^{2}+i0\right)\left(\ell_{4}^{2}-m_{2}^{2}+i0\right)}\right]\nonumber 
\end{align}
where we introduced shorthand notations 
\begin{align}
\bar{\Pi}_{M_{1}}^{h_{1},h_{2}} & =\bar{v}_{h_{2}}\left(\tilde{\ell}_{2}\right)\gamma_{+}\Pi_{M_{1}}^{SS_{Z}}\left(p_{1},\,0\right)\gamma^{+}u_{h_{1}}\left(\tilde{\ell}_{1}\right),\label{eq:P1}\\
\bar{\Pi}_{M_{2}}^{h_{3},h_{4}} & =\bar{v}_{h_{4}}\left(\tilde{\ell}_{4}\right)\gamma_{+}\Pi_{M_{2}}^{SS_{Z}}\left(p_{2},\,0\right)\gamma^{+}u_{h_{3}}\left(\tilde{\ell}_{3}\right).\label{eq:P2}
\end{align}
As we will show now, the matrices $\bar{\Pi}_{M_{1}}^{h_{1},h_{2}},\,\bar{\Pi}_{M_{2}}^{h_{3},h_{4}}$
do not depend on the integration momenta $\ell_{n}$. In order to
demonstrate this, we will rewrite the explicit form of the NRQCD projectors~(\ref{eq:NRQCD_Projector})
as
\begin{equation}
\Pi_{M}^{SS_{Z}}\left(p_{M},\,0\right)=\left(\frac{\hat{p}_{M}}{2}-m\right)\Gamma_{M}\left(\frac{\hat{p}_{M}}{2}+m\right)=\sum_{h,\bar{h}}v_{\bar{h}}\left(\frac{p_{M}}{2}\right)\bar{v}_{\bar{h}}\left(\frac{p_{M}}{2}\right)\Gamma_{M}u_{h}\left(\frac{p_{M}}{2}\right)\bar{u}_{h}\left(\frac{p_{M}}{2}\right),\label{eq:PP}
\end{equation}
where $\Gamma_{M}$ is the appropriate (spin/helicity dependent) Dirac
matrix multiplied by the corresponding LDME. Substituting~(\ref{eq:PP})
into~(\ref{eq:P1},\ref{eq:P2}), we may obtain for the NRQCD projectors
in helicity basis
\begin{align}
\bar{\Pi}_{M}^{h_{1},h_{2}} & =\bar{v}_{h_{2}}\left(\tilde{\ell}_{2}\right)\gamma_{+}\left(\sum_{h,\bar{h}}v_{\bar{h}}\left(\frac{p_{M}}{2}\right)\bar{v}_{\bar{h}}\left(\frac{p_{M}}{2}\right)\Gamma_{M}u_{h}\left(\frac{p_{M}}{2}\right)\bar{u}_{h}\left(\frac{p_{M}}{2}\right)\right)\gamma^{+}u_{h_{1}}\left(\tilde{\ell}_{1}\right)=\left(p_{M}^{+}\right)^{2}\bar{v}_{h_{2}}\left(\frac{p_{M}}{2}\right)\Gamma_{M}u_{h_{1}}\left(\frac{p_{M}}{2}\right),
\end{align}
where we used identities for the light-cone spinors $\bar{v}_{h_{1}}(p)\gamma_{+}v_{h_{2}}(q)=\bar{u}_{h_{1}}(p)\gamma_{+}u_{h_{2}}(q)=2\delta_{h_{1},h_{2}}\sqrt{p^{+}q^{+}}$~\cite{Lepage:1980fj},
as well as $\tilde{\ell}_{1}^{+}=\tilde{\ell}_{2}^{+}=p_{M}^{+}/2$,
which is a consequence of the fact that shockwave does not change
the ``+'' momentum of the parton. The structure of~(\ref{eq:R4-1},~\ref{eq:R4Prime-1})
demonstrates that the impact factors essentially reduce to a convolution
of the light-cone wave functions with NRQCD projectors in helicity
basis. Indeed, if we express $\ell_{2}$ in terms of the other light-cone
momenta and take the integrals over $\ell_{1}^{-},\ell_{3}^{-}$ using
identities based on the Cauchy theorem
\begin{align}
 & \int_{-\infty}^{+\infty}\frac{d\ell_{1}^{-}}{2\pi}\frac{1}{\left(2\ell_{1}^{+}\ell{}_{1}^{-}-\boldsymbol{\ell}_{1,\perp}^{2}-m_{1}^{2}+i0\right)\left(2\left(k^{+}+\ell_{1}^{+}\right)\left(k^{-}+\ell_{1}^{-}\right)-\left(\boldsymbol{k}_{\perp}+\boldsymbol{\ell}_{1,\perp}\right)^{2}-m_{1}^{2}+i0\right)}\times\label{eq:Cauchy_1}\\
 & \times\frac{1}{\left(2\left(q^{+}-k^{+}-\ell_{1}^{+}\right)\left(q^{-}-k^{-}-\ell_{1}^{-}\right)-\left(\boldsymbol{k}+\boldsymbol{\ell}_{1,\perp}\right)^{2}-m_{1}^{2}+i0\right)}=\nonumber \\
 & =\frac{i\Theta\left(q^{+}-k^{+}-\ell_{1}^{+}\right)}{8\ell_{1}^{+}\left(\ell_{1}^{+}+k^{+}\right)\left(q^{+}-k^{+}-\ell_{1}^{+}\right)\left[-\frac{q^{+}\left(\boldsymbol{\ell}_{1,\perp}^{2}+m_{1}^{2}\right)}{2\left(q^{+}-k^{+}-\ell_{1}^{+}\right)\left(k^{+}+\ell_{1}^{+}\right)}\right]\left[-\frac{\boldsymbol{k}_{\perp}^{2}}{2k^{+}}-\frac{\boldsymbol{\ell}_{1\perp}^{2}+m_{1}^{2}}{2\ell_{0}^{+}}-\frac{\left(\boldsymbol{k}_{\perp}+\boldsymbol{\ell}_{1\perp}\right)^{2}+m_{1}^{2}}{2\left(q^{+}-k^{+}-\ell_{1}^{+}\right)}\right]},\nonumber 
\end{align}
\begin{align}
 & \int_{-\infty}^{+\infty}\frac{d\ell_{1}^{-}}{2\pi}\frac{\left(k^{-}+\ell_{1}^{-}\right)}{\left(2\ell_{1}^{+}\ell{}_{1}^{-}-\boldsymbol{\ell}_{1\perp}^{2}-m_{1}^{2}+i0\right)\left(2\left(k^{+}+\ell_{1}^{+}\right)\left(k^{-}+\ell_{1}^{-}\right)-\left(\boldsymbol{k}_{\perp}+\boldsymbol{\ell}_{1\perp}\right)^{2}-m_{1}^{2}+i0\right)}\times\label{eq:Cauchy_1-1}\\
 & \times\frac{1}{\left(2\left(q^{+}-k^{+}-\ell_{1}^{+}\right)\left(q^{-}-k^{-}-\ell_{1}^{-}\right)-\left(\boldsymbol{k}_{\perp}+\boldsymbol{\ell}_{1\perp}\right)^{2}-m_{1}^{2}+i0\right)}=\nonumber \\
 & =\frac{i\Theta\left(q^{+}-k^{+}-\ell_{1}^{+}\right)\frac{\left(\boldsymbol{k}_{\perp}+\boldsymbol{\ell}_{1\perp}\right)^{2}+m_{1}^{2}}{2\left(k^{+}+\ell_{1}^{+}\right)}}{8\ell_{1}^{+}\left(\ell_{1}^{+}+k^{+}\right)\left(q^{+}-k^{+}-\ell_{1}^{+}\right)\left[q^{-}-\frac{q^{+}\left(\boldsymbol{\ell}_{1\perp}^{2}+m_{1}^{2}\right)}{2\left(q^{+}-k^{+}-\ell_{1}^{+}\right)\left(k^{+}+\ell_{1}^{+}\right)}\right]\left[q^{-}-\frac{\boldsymbol{k}_{\perp}^{2}}{2k^{+}}-\frac{\boldsymbol{\ell}_{1\perp}^{2}+m_{1}^{2}}{2\ell_{1}^{+}}-\frac{\left(\boldsymbol{k}_{\perp}+\boldsymbol{\ell}_{1\perp}\right)^{2}+m_{1}^{2}}{2\left(q^{+}-k^{+}-\ell_{1}^{+}\right)}\right]}+\nonumber \\
 & +\frac{i\Theta\left(q^{+}-k^{+}-\ell_{1}^{+}\right)}{8\ell_{1}^{+}\left(\ell_{1}^{+}+k^{+}\right)\left(q^{+}-k^{+}-\ell_{1}^{+}\right)\left[q^{-}-\frac{\boldsymbol{k}_{\perp}^{2}}{2k^{+}}-\frac{\boldsymbol{\ell}_{1\perp}^{2}+m_{1}^{2}}{2\ell_{1}^{+}}-\frac{\left(\boldsymbol{k}_{\perp}+\boldsymbol{\ell}_{1\perp}\right)^{2}+m_{1}^{2}}{2\left(q^{+}-k^{+}-\ell_{1}^{+}\right)}\right]},\nonumber 
\end{align}
\begin{align}
 & \int_{-\infty}^{+\infty}\frac{d\ell_{3}^{-}}{2\pi}\frac{1}{\left(2\ell_{3}^{+}\ell_{3}^{-}-\boldsymbol{\ell}{}_{3\perp}^{2}-m_{2}^{2}+i0\right)\left(2\left(k^{+}-\ell_{3}^{+}\right)\left(k^{-}-\ell_{3}^{-}\right)-\left(\boldsymbol{k}_{\perp}-\boldsymbol{\ell}_{3\perp}\right)^{2}-m_{2}^{2}+i0\right)}=\label{eq:Cauchy_2}\\
 & =\frac{\Theta\left(k^{+}-\ell_{3}^{+}\right)}{4\ell_{3}^{+}\left(k^{+}-\ell_{3}^{+}\right)}\,\frac{i}{k^{-}-\frac{\boldsymbol{\ell}_{3\perp}^{2}+m^{2}}{2\ell_{3}^{+}}-\frac{\left(\boldsymbol{k}_{\perp}-\boldsymbol{\ell}_{3\perp}\right)^{2}+m_{2}^{2}}{2\left(k^{+}-\ell_{3}^{+}\right)}},\nonumber 
\end{align}
where $k=\ell_{3}+\ell_{4}$, we recover the light-cone energy denominators
which should appear in diagram $4$ according to the light-cone perturbation
theory suggested long ago in~\cite{Lepage:1980fj,Brodsky:1997de}.
Similarly, if we express $\ell_{1}$ in terms of $\ell_{2},\,k$ and
apply identities~(\ref{eq:Cauchy_1},\ref{eq:Cauchy_1-1}) with permuted
dummy integration variable $\ell_{1}^{-}\to\ell_{2}^{-}$ , we will
recover the light-cone energy denominators for the diagram $4'$.
The remaining integration over the transverse components is straightforward
and allows to express the impact factors as convolutions

\begin{align}
R_{4}\left(\boldsymbol{x}_{1},\,\boldsymbol{x}_{2},\,\boldsymbol{x}_{3},\,\boldsymbol{x}_{4}\right) & =\sum_{h_{1},...,h_{4}}\bar{\Pi}_{M_{1}}^{h_{1},h_{2}}\bar{\Pi}_{M_{2}}^{h_{3},h_{4}}A_{h_{1},h_{2},h_{3},h_{4}}\left(\left\{ \alpha_{i},\,\boldsymbol{x}_{i}\right\} \right)e^{-i\boldsymbol{p}_{1}\cdot\left(\boldsymbol{x}_{1}+\boldsymbol{x}_{2}\right)/2}e^{-i\boldsymbol{p}_{2}\cdot\left(\boldsymbol{x}_{3}+\boldsymbol{x}_{4}\right)/2}\label{eq:R4-1-1}
\end{align}
\begin{align}
R_{4'}\left(\boldsymbol{x}_{1},\,\boldsymbol{x}_{2},\,\boldsymbol{x}_{3},\,\boldsymbol{x}_{4}\right) & =\sum_{h_{1},...,h_{4}}\bar{\Pi}_{M_{1}}^{h_{1},h_{2}}\bar{\Pi}_{M_{2}}^{h_{3},h_{4}}B_{h_{1},h_{2},h_{3},h_{4}}\left(\left\{ \alpha_{i},\,\boldsymbol{x}_{i}\right\} \right)e^{-i\boldsymbol{p}_{1}\cdot\left(\boldsymbol{x}_{1}+\boldsymbol{x}_{2}\right)/2}e^{-i\boldsymbol{p}_{2}\cdot\left(\boldsymbol{x}_{3}+\boldsymbol{x}_{4}\right)/2}\times\label{eq:R4Prime-1-1}
\end{align}

where the functions $A_{h_{1},h_{2},h_{3},h_{4}},\,B_{h_{1},h_{2},h_{3},h_{4}}$
are defined as
\begin{align}
A_{h_{1},h_{2},h_{3},h_{4}} & \left(\left\{ \alpha_{i},\,\boldsymbol{x}_{i}\right\} \right)=e^{i\boldsymbol{p}_{1}\cdot\left(\boldsymbol{x}_{1}+\boldsymbol{x}_{2}\right)/2}e^{i\boldsymbol{p}_{2}\cdot\left(\boldsymbol{x}_{3}+\boldsymbol{x}_{4}\right)/2}\sum_{h_{1},...,h_{4}}\int\prod_{n=1}^{4}\left(\frac{d^{3}\ell_{n}}{\left(2\pi\right)^{3}}e^{i\boldsymbol{\ell}_{n}\cdot\boldsymbol{x}_{n}}\right)\left(2\pi\right)^{3}\delta^{3}\left(q-\sum_{n}\ell_{n}\right)\times\label{eq:R4-1-2}\\
 & \times\left(ig\right)^{3}\left[\frac{\bar{u}_{h_{1}}\left(\tilde{\ell}_{1}\right)\gamma^{\mu}S\left(\ell_{1}-\ell_{3}-\ell_{4}\right)\hat{\varepsilon}\left(q\right)v_{h_{2}}\left(\tilde{\ell}_{2}\right)}{\left(\ell_{1}^{2}-m_{1}^{2}+i0\right)\left(\ell_{2}^{2}-m_{1}^{2}+i0\right)}\right]{\rm \Pi_{\mu\nu}\left(\ell_{3}+\ell_{4}\right)}\left[\frac{\bar{u}_{h_{3}}\left(\tilde{\ell}_{3}\right)\gamma^{\nu}v_{h_{4}}\left(\tilde{\ell}_{4}\right)}{\left(\ell_{3}^{2}-m_{2}^{2}+i0\right)\left(\ell_{4}^{2}-m_{2}^{2}+i0\right)}\right]\nonumber 
\end{align}

\begin{align}
B_{h_{1},h_{2},h_{3},h_{4}} & \left(\left\{ \alpha_{i},\,\boldsymbol{x}_{i}\right\} \right)=e^{i\boldsymbol{p}_{1}\cdot\left(\boldsymbol{x}_{1}+\boldsymbol{x}_{2}\right)/2}e^{i\boldsymbol{p}_{2}\cdot\left(\boldsymbol{x}_{3}+\boldsymbol{x}_{4}\right)/2}\int\prod_{n=1}^{4}\left(\frac{d^{3}\ell_{n}}{\left(2\pi\right)^{3}}e^{i\boldsymbol{\ell}_{n}\cdot\boldsymbol{x}_{n}}\right)\left(2\pi\right)^{3}\delta^{3}\left(q-\sum_{n}\ell_{n}\right)\times\label{eq:R4Prime-1-2}\\
 & \times\left(ig\right)^{3}\left[\frac{\bar{u}_{h_{1}}\left(\tilde{\ell}_{1}\right)\hat{\varepsilon}\left(q\right)S\left(\ell_{1}-q\right)\gamma^{\mu}v_{h_{2}}\left(\tilde{\ell}_{2}\right)}{\left(\ell_{1}^{2}-m_{1}^{2}+i0\right)\left(\ell_{2}^{2}-m_{1}^{2}+i0\right)}\right]\Pi_{\mu\nu}\left(\ell_{3}+\ell_{4}\right)\left[\frac{\bar{u}_{h_{3}}\left(\tilde{\ell}_{3}\right)\gamma^{\nu}v_{h_{4}}\left(\tilde{\ell}_{4}\right)}{\left(\ell_{3}^{2}-m_{2}^{2}+i0\right)\left(\ell_{4}^{2}-m_{2}^{2}+i0\right)}\right]\nonumber \\
 & =-A_{h_{2},h_{1},h_{4},h_{3}}\left(\alpha_{2},\,\boldsymbol{x}_{2},\,\alpha_{1},\,\boldsymbol{x}_{1},\,\alpha_{4},\,\boldsymbol{x}_{4},\,\alpha_{3},\,\boldsymbol{x}_{3}\right)\nonumber 
\end{align}
and the corresponding light-cone fractions $\alpha_{i}$ in the heavy
quark mass limit are given by 
\begin{equation}
\alpha_{1}=\alpha_{2}=\frac{p_{M_{1}}^{+}}{2q^{+}},\quad\alpha_{3}=\alpha_{4}=\frac{p_{M_{1}}^{+}}{2q^{+}}.
\end{equation}
The evaluation of the expressions for $A_{h_{1},h_{2},h_{3},h_{4}},\,B_{h_{1},h_{2},h_{3},h_{4}}$
essentially coincides with the light-cone wave functions of the $\bar{Q}Q\bar{Q}Q$
Fock state in the photon (namely, the partonic amplitude $\gamma\to\bar{Q}Q\bar{Q}Q$),
which was done in~\cite{Andrade:2022rbn}, and after several algebraic
simplifications may be reduced to the form 

\begin{align}
A\left(\left\{ \alpha_{i},\,\boldsymbol{r}_{i}\right\} \right) & =-\frac{2\sqrt{4\pi\alpha_{s}\left(\mu\right)}\alpha_{s}\left(\mu\right)\,}{\pi^{3}\left(1-\alpha_{1}-\alpha_{2}\right)^{2}\sqrt{\alpha_{1}\alpha_{2}}\,}\int\frac{q_{1}dq_{1}\,k_{2}dk_{2}}{\frac{\bar{\alpha}_{2}q_{1}^{2}}{\alpha_{1}\left(1-\alpha_{1}-\alpha_{2}\right)}+\frac{m_{1}^{2}\left(\alpha_{1}+\alpha_{2}\right)}{\alpha_{1}\alpha_{2}}+\frac{k_{2}^{2}}{\alpha_{2}\bar{\alpha}_{2}}}\times\label{eq:Amp_p_explicit-4-4-1}\\
 & \times\frac{1}{k_{2}^{2}+m_{1}^{2}}\sqrt{\frac{\alpha_{2}}{\alpha_{1}}}\left[\left(\alpha_{2}\delta_{\gamma,a_{2}}-\bar{\alpha}_{2}\delta_{\gamma,-a_{2}}\right)\left(\bar{\alpha}_{2}\delta_{\lambda,\,a_{1}}+\alpha_{1}\delta_{\lambda,-a_{1}}\right)\delta_{a_{1},-a_{2}}\times\right.\nonumber \\
 & \times\left(\boldsymbol{n}_{2,134}\cdot\boldsymbol{\varepsilon}_{\gamma}\right)\left(\boldsymbol{n}_{1,34}\cdot\boldsymbol{\varepsilon}_{\lambda}^{*}\right)k_{2}\,J_{1}\left(k_{2}\left|\boldsymbol{x}_{2}-\boldsymbol{b}_{134}\right|\right)q_{1}J_{1}\left(q_{1}\left|\boldsymbol{x}_{1}-\boldsymbol{b}_{34}\right|\right)+\nonumber \\
 & +\frac{m_{q}^{2}}{2}\,\delta_{\lambda,-a_{1}}\delta_{\gamma,a_{2}}\delta_{a_{1},-a_{2}}J_{0}\left(k_{2}\left|\boldsymbol{x}_{2}-\boldsymbol{b}_{134}\right|\right)J_{0}\left(q_{1}\left|\boldsymbol{x}_{1}-\boldsymbol{b}_{34}\right|\right)\frac{\left(1-\alpha_{1}-\alpha_{2}\right)^{2}}{1-\alpha_{2}}\nonumber \\
 & -\frac{im_{q}}{\sqrt{2}}\,{\rm sign}\left(a_{2}\right)\delta_{\gamma,a_{2}}\delta_{a_{1},a_{2}}\left(\bar{\alpha}_{2}\delta_{\lambda,\,a_{1}}+\alpha_{1}\delta_{\lambda,-a_{1}}\right)\times\nonumber \\
 & \times\boldsymbol{n}_{1,34}\cdot\boldsymbol{\varepsilon}_{\lambda}^{*}q_{1}J_{1}\left(q_{1}\left|\boldsymbol{x}_{1}-\boldsymbol{b}_{34}\right|\right)J_{0}\left(k_{2}\left|\boldsymbol{x}_{2}-\boldsymbol{b}_{134}\right|\right)\nonumber \\
 & -\frac{im_{q}}{\sqrt{2}}{\rm sign}\left(a_{1}\right)\delta_{\lambda,-a_{1}}\left(\alpha_{2}\delta_{\gamma,a_{2}}-\bar{\alpha}_{2}\delta_{\gamma,-a_{2}}\right)\delta_{a_{1},a_{2}}\frac{\left(1-\alpha_{1}-\alpha_{2}\right)^{2}}{1-\alpha_{2}}\times\nonumber \\
 & \times\left.\left(\boldsymbol{n}_{2,134}\cdot\boldsymbol{\varepsilon}_{\gamma}\right)k_{2}\,J_{1}\left(k_{2}\left|\boldsymbol{x}_{2}-\boldsymbol{b}_{134}\right|\right)J_{0}\left(q_{1}\left|\boldsymbol{x}_{1}-\boldsymbol{b}_{34}\right|\right)\right]\times\nonumber \\
 & \times\Psi_{a_{3},a_{4}}^{-\lambda}\left(\frac{\alpha_{3}}{\alpha_{3}+\alpha_{4}},\,\boldsymbol{r}_{34},\,m_{2},\,\sqrt{m_{2}^{2}+\frac{\alpha_{3}\alpha_{4}}{\alpha_{3}+\alpha_{4}}\left[\frac{\bar{\alpha}_{2}q_{1}^{2}}{\alpha_{1}\left(1-\alpha_{1}-\alpha_{2}\right)}+\frac{m_{1}^{2}\left(\alpha_{1}+\alpha_{2}\right)}{\alpha_{1}\alpha_{2}}+\frac{k_{2}^{2}}{\alpha_{2}\bar{\alpha}_{2}}\right]}\right),\nonumber 
\end{align}
where
\begin{equation}
\Psi_{h\bar{h}}^{\lambda}\left(z,\,\boldsymbol{r}_{12},\,m_{q},a\right)=-\frac{2}{(2\pi)}\left[\left(z\delta_{\lambda,h}-(1-z)\delta_{\lambda,-h}\right)\delta_{h,-\bar{h}}i\boldsymbol{\varepsilon}_{\lambda}\cdot\nabla-\frac{m_{q}}{\sqrt{2}}\,{\rm sign}(h)\delta_{\lambda,h}\delta_{h,\bar{h}}\right]K_{0}\left(a\,\boldsymbol{r}\right).\label{eq:PsiSplitting-1-1}
\end{equation}
For the impact factor of $R_{8}$ we may repeat the same analysis
and replace propagators of the partons which cross the shockwave with~(\ref{eq:onshell}).
After some algebraic simplifications we may get a structure similar
to~(\ref{eq:R4-1-1},~\ref{eq:R4Prime-1-1}), namely 

\begin{align}
R_{8}\left(\boldsymbol{x}_{1},\,\boldsymbol{x}_{2},\,\boldsymbol{x}_{3},\,\boldsymbol{x}_{4}\right) & =\sum_{h_{1},...,h_{4}}\bar{\Pi}_{M_{1}}^{h_{1},h_{2}}\bar{\Pi}_{M_{2}}^{h_{3},h_{4}}C_{h_{1},h_{2},h_{3},h_{4}}\left(\left\{ \alpha_{i},\,\boldsymbol{x}_{i}\right\} \right)e^{-i\boldsymbol{p}_{1}\cdot\left(\boldsymbol{x}_{1}+\boldsymbol{x}_{2}\right)/2}e^{-i\boldsymbol{p}_{2}\cdot\left(\boldsymbol{x}_{3}+\boldsymbol{x}_{4}\right)/2}\label{eq:R4-1-1-1}
\end{align}
where the function $C_{h_{1},h_{2},h_{3},h_{4}}$ is defined as
\begin{align}
C_{h_{1},h_{2},h_{3},h_{4}} & \left(\left\{ \alpha_{i},\,\boldsymbol{x}_{i}\right\} \right)=e^{i\boldsymbol{p}_{1}\cdot\left(\boldsymbol{x}_{1}+\boldsymbol{x}_{2}\right)/2}e^{i\boldsymbol{p}_{2}\cdot\left(\boldsymbol{x}_{3}+\boldsymbol{x}_{4}\right)/2}\left(ig\right)^{3}\int\prod_{n=1}^{4}\left(\frac{d^{3}\ell_{n}}{\left(2\pi\right)^{3}}e^{i\boldsymbol{\ell}_{n}\cdot\boldsymbol{x}_{n}}\right)\varepsilon^{\lambda}(q)V_{\lambda\alpha\beta}\times\label{eq:R8-1}\\
 & \times\Pi_{\alpha\mu}\left(\ell_{1}+\ell_{2}\right)\left[\frac{\bar{u}_{h_{1}}\left(\tilde{\ell}_{1}\right)\gamma^{\mu}v_{h_{2}}\left(\tilde{\ell}_{2}\right)}{\left(\ell_{1}^{2}-m_{1}^{2}+i0\right)\left(\ell_{2}^{2}-m_{1}^{2}+i0\right)}\right]{\rm \Pi_{\beta\nu}\left(\ell_{3}+\ell_{4}\right)}\left[\frac{\bar{u}_{h_{3}}\left(\tilde{\ell}_{3}\right)\gamma^{\nu}v_{h_{4}}\left(\tilde{\ell}_{4}\right)}{\left(\ell_{3}^{2}-m_{2}^{2}+i0\right)\left(\ell_{4}^{2}-m_{2}^{2}+i0\right)}\right]\times\nonumber \\
 & \times\left(2\pi\right)^{3}\delta^{3}\left(q-\sum_{n}\ell_{n}\right)\nonumber 
\end{align}
and is antisymmetric w.r.t. permutation of pairs of the final state
quarks $(12)\leftrightarrow(34)$. The expressions in the square brackets
of the second line of~(\ref{eq:R8-1}) clearly coincide with the
wave functions of $\gamma\to\bar{Q}Q$ in the momentum space~\cite{Lappi:2016oup},
for this reason the function $C_{h_{1},h_{2},h_{3},h_{4}}\left(\left\{ \alpha_{i},\,\boldsymbol{x}_{i}\right\} \right)$
in light-cone representation can be represented as a product of the
wave functions $\Psi_{h_{1},h_{2}}^{\lambda}\left(\boldsymbol{r}_{12},\,m_{1}\right)$,
$\Psi_{h_{3},h_{4}}^{\lambda}\left(\boldsymbol{r}_{34},\,m_{2}\right)$
defined in~(\ref{eq:PsiSplitting-1-1}) convoluted with gluon propagators.
Summarizing our findings, we may rewrite the amplitude of the $g\to M_{1}M_{2}X$
subprocess as 
\begin{align}
 & \mathcal{A}_{a}\left(p_{1},\,S_{1};\,p_{2},\,S_{2}\right)=\bar{\Pi}_{M_{1}}^{h_{1},h_{2}}\bar{\Pi}_{M_{2}}^{h_{3},h_{4}}\int\prod_{n=1}^{4}\left(d^{2}\boldsymbol{x}_{n}\right)e^{-i\boldsymbol{p}_{1}\cdot\left(\boldsymbol{x}_{1}+\boldsymbol{x}_{2}\right)/2}e^{-i\boldsymbol{p}_{2}\cdot\left(\boldsymbol{x}_{3}+\boldsymbol{x}_{4}\right)/2}\times\label{eq:Amp}\\
 & \times\left[\bar{A}_{h_{1},h_{2},h_{3},h_{4}}\left(\left\{ \alpha_{i},\,\boldsymbol{x}_{i}\right\} \right)T_{a}^{(4)}\left(\boldsymbol{x}_{1},\,\boldsymbol{x}_{2},\,\boldsymbol{x}_{3},\,\boldsymbol{x}_{4}\right)+\bar{B}_{h_{1},h_{2},h_{3},h_{4}}\left(\left\{ \alpha_{i},\,\boldsymbol{x}_{i}\right\} \right)T_{a}^{(4')}\left(\boldsymbol{x}_{1},\,\boldsymbol{x}_{2},\,\boldsymbol{x}_{3},\,\boldsymbol{x}_{4}\right)\right]\nonumber 
\end{align}
where we defined shorthand notations
\begin{align}
 & \bar{A}_{a_{1},a_{2},a_{3},a_{4}}\left(\left\{ \alpha_{i},\,\boldsymbol{x}_{i}\right\} \right)=A_{a_{1},a_{2},a_{3},a_{4}}\left(\left\{ \alpha_{i},\,\boldsymbol{x}_{i}\right\} \right)-C_{a_{1},a_{2},a_{3},a_{4}}\left(\left\{ \alpha_{i},\,\boldsymbol{x}_{i}\right\} \right),\\
 & \bar{B}_{a_{1},a_{2},a_{3},a_{4}}\left(\left\{ \alpha_{i},\,\boldsymbol{x}_{i}\right\} \right)=B_{a_{1},a_{2},a_{3},a_{4}}\left(\left\{ \alpha_{i},\,\boldsymbol{x}_{i}\right\} \right)+C_{a_{1},a_{2},a_{3},a_{4}}\left(\left\{ \alpha_{i},\,\boldsymbol{x}_{i}\right\} \right),
\end{align}
and used the last two identities in~(\ref{eq:T8}) for simplifications.
In the inclusive cross-section we have to average (integrate over
the color sources) the square of the amplitude $\left|\mathcal{A}_{a}\right|^{2}$,
which requires evaluation of the convolutions of different elements
$S_{2n}^{(a)}$; for the latter we can use an identity 
\begin{align}
S_{2n}^{(b)}\left(\boldsymbol{x}_{1},\,...,\boldsymbol{x}_{2n}\right)\,S_{2m}^{(b)}\left(\boldsymbol{y}_{1},\,...,\,\boldsymbol{y}_{2m}\right) & =\frac{1}{2N_{c}}\left\langle S_{2n+2m}\left(\boldsymbol{x}_{1},\,...,\boldsymbol{x}_{2n},\boldsymbol{y}_{1},\,...,\,\boldsymbol{y}_{2m}\right)\right.-\\
 & -\left.S_{2n}\left(\boldsymbol{x}_{1},\,...,\boldsymbol{x}_{2n}\right)\,S_{2m}\left(\boldsymbol{y}_{1},\,...,\,\boldsymbol{y}_{2m}\right)\right\rangle \nonumber 
\end{align}
which also follows from~(\ref{eq:Fierz}). The final result of this
procedure is
\begin{align}
\frac{d\sigma\left(g+p\to M_{1}+M_{2}+X\right)}{dy_{1}d^{2}\boldsymbol{p}_{1}^{\perp}dy_{2}d^{2}\boldsymbol{p}_{2}^{\perp}} & =\frac{1}{16\pi}\left|\mathcal{A}_{a}\left(p_{M_{1}},\,S_{1};\,p_{M_{2}},\,S_{2}\right)\right|^{2}=\frac{N_{c}^{2}}{64\pi}\int\prod_{n=1}^{4}\left(d^{2}\boldsymbol{x}_{n}d^{2}\boldsymbol{y}_{n}\right)\times\label{eq:ASq}\\
 & \times e^{-i\boldsymbol{p}_{1}\cdot\left(\boldsymbol{x}_{1}+\boldsymbol{x}_{2}-\boldsymbol{y}_{1}-\boldsymbol{y}_{2}\right)/2}e^{-i\boldsymbol{p}_{2}\cdot\left(\boldsymbol{x}_{3}+\boldsymbol{x}_{4}-\boldsymbol{y}_{3}-\boldsymbol{y}_{4}\right)/2}\bar{\Pi}_{M_{1}}^{a_{1},a_{2}}\bar{\Pi}_{M_{2}}^{a_{3},a_{4}}\bar{\Pi}_{M_{1}}^{b_{1},b_{2}*}\bar{\Pi}_{M_{2}}^{b_{3},b_{4}*}\times\nonumber \\
 & \times\left[S_{4}\left(\boldsymbol{x}_{1},\,\boldsymbol{x}_{2},\,\boldsymbol{y}_{2},\,\boldsymbol{y}_{1}\right)S_{2}\left(\boldsymbol{x}_{4},\,\boldsymbol{x}_{3}\right)S_{2}\left(\boldsymbol{y}_{4},\,\boldsymbol{y}_{3}\right)\psi_{a_{1},a_{2},a_{3},a_{4}}\left(\left\{ \alpha_{i},\,\boldsymbol{x}_{i}\right\} \right)\psi_{b_{1},b_{2},b_{3},b_{4}}\left(\left\{ \alpha_{i},\,\boldsymbol{y}_{i}\right\} \right)^{*}\right.\nonumber \\
 & -\psi_{b_{1},b_{2},b_{3},b_{4}}\left(\left\{ \alpha_{i},\,\boldsymbol{y}_{i}\right\} \right)^{*}S_{2}\left(\boldsymbol{y}_{4},\,\boldsymbol{y}_{3}\right)\times\nonumber \\
 & \times\left[\bar{A}_{a_{1},a_{2},a_{3},a_{4}}S_{6}\left(\boldsymbol{x}_{3},\,\boldsymbol{x}_{4},\,\boldsymbol{x}_{1},\,\boldsymbol{x}_{2},\,\boldsymbol{y}_{2},\,\boldsymbol{y}_{1}\right)+\bar{B}_{a_{1},a_{2},a_{3},a_{4}}S_{6}\left(\boldsymbol{x}_{1},\,\boldsymbol{x}_{2},\,\boldsymbol{x}_{3},\,\boldsymbol{x}_{4},\,\boldsymbol{y}_{2},\,\boldsymbol{y}_{1}\right)\right]\nonumber \\
 & -\psi_{a_{1},a_{2},a_{3},a_{4}}\left(\left\{ \alpha_{i},\,\boldsymbol{x}_{i}\right\} \right)S_{2}\left(\boldsymbol{x}_{4},\,\boldsymbol{x}_{3}\right)\times\nonumber \\
 & \times\left[\bar{A}_{b_{1},b_{2},b_{3},b_{4}}^{*}S_{6}\left(\boldsymbol{y}_{3},\,\boldsymbol{y}_{4},\,\boldsymbol{y}_{1},\,\boldsymbol{y}_{2},\,\boldsymbol{x}_{2},\,\boldsymbol{x}_{1}\right)+\bar{B}_{b_{1},b_{2},b_{3},b_{4}}^{*}S_{6}\left(\boldsymbol{y}_{4},\,\boldsymbol{y}_{3},\,\boldsymbol{y}_{2},\,\boldsymbol{y}_{1},\,\boldsymbol{x}_{2},\,\boldsymbol{x}_{1}\right)\right]\nonumber \\
 & \qquad+\bar{A}_{a_{1},a_{2},a_{3},a_{4}}\bar{A}_{b_{1},b_{2},b_{3},b_{4}}^{*}S_{8}\left(\boldsymbol{x}_{3},\,\boldsymbol{x}_{4},\,\boldsymbol{x}_{1},\,\boldsymbol{x}_{2},\,\boldsymbol{y}_{2},\,\boldsymbol{y}_{1},\,\boldsymbol{y}_{4},\,\boldsymbol{y}_{3}\right)\nonumber \\
 & \qquad+\bar{B}_{a_{1},a_{2},a_{3},a_{4}}\bar{A}_{b_{1},b_{2},b_{3},b_{4}}^{*}S_{8}\left(\boldsymbol{x}_{1},\,\boldsymbol{x}_{2},\,\boldsymbol{x}_{3},\,\boldsymbol{x}_{4},\,\boldsymbol{y}_{2},\,\boldsymbol{y}_{1},\,\boldsymbol{y}_{4},\,\boldsymbol{y}_{3}\right)\nonumber \\
 & \qquad+\bar{A}_{a_{1},a_{2},a_{3},a_{4}}\bar{B}_{b_{1},b_{2},b_{3},b_{4}}^{*}S_{8}\left(\boldsymbol{x}_{3},\,\boldsymbol{x}_{4},\,\boldsymbol{x}_{1},\,\boldsymbol{x}_{2},\,\boldsymbol{y}_{4},\,\boldsymbol{y}_{3},\,\boldsymbol{y}_{2},\,\boldsymbol{y}_{1}\right)+\nonumber \\
 & \quad\left.+\bar{B}_{a_{1},a_{2},a_{3},a_{4}}\bar{B}_{b_{1},b_{2},b_{3},b_{4}}^{*}S_{8}\left(\boldsymbol{x}_{1},\,\boldsymbol{x}_{2},\,\boldsymbol{x}_{3},\,\boldsymbol{x}_{4},\,\boldsymbol{y}_{4},\,\boldsymbol{y}_{3},\,\boldsymbol{y}_{2},\,\boldsymbol{y}_{1}\right)\right]\nonumber 
\end{align}

where we introduced a shorthand notation 
\[
\psi\left(\left\{ \alpha_{i},\,\boldsymbol{x}_{i}\right\} \right)=A\left(\left\{ \alpha_{i},\,\boldsymbol{x}_{i}\right\} \right)+B\left(\left\{ \alpha_{i},\,\boldsymbol{x}_{i}\right\} \right)=\bar{A}\left(\left\{ \alpha_{i},\,\boldsymbol{x}_{i}\right\} \right)+\bar{B}\left(\left\{ \alpha_{i},\,\boldsymbol{x}_{i}\right\} \right)
\]

and contracted color indices of incoming gluons using identities
\begin{align}
 & T_{a}^{(4)}\left(\boldsymbol{x}_{1},\,\boldsymbol{x}_{2},\,\boldsymbol{x}_{3},\,\boldsymbol{x}_{4}\right)\left(T_{a}^{(4)}\left(\boldsymbol{y}_{1},\,\boldsymbol{y}_{2},\,\boldsymbol{y}_{3},\,\boldsymbol{y}_{4}\right)\right)^{*}=\nonumber \\
 & \frac{N_{c}^{2}}{4}\left[S_{8}\left(\boldsymbol{x}_{3},\,\boldsymbol{x}_{4},\,\boldsymbol{x}_{1},\,\boldsymbol{x}_{2},\,\boldsymbol{y}_{2},\,\boldsymbol{y}_{1},\,\boldsymbol{y}_{4},\,\boldsymbol{y}_{3}\right)+S_{4}\left(\boldsymbol{x}_{1},\,\boldsymbol{x}_{2},\,\boldsymbol{y}_{2},\,\boldsymbol{y}_{1}\right)S_{2}\left(\boldsymbol{x}_{4},\,\boldsymbol{x}_{3}\right)S_{2}\left(\boldsymbol{y}_{4},\,\boldsymbol{y}_{3}\right)-\right.\\
 & -\left.S_{6}\left(\boldsymbol{x}_{3},\,\boldsymbol{x}_{4},\,\boldsymbol{x}_{1},\,\boldsymbol{x}_{2},\,\boldsymbol{y}_{2},\,\boldsymbol{y}_{1}\right)S_{2}\left(\boldsymbol{y}_{4},\,\boldsymbol{y}_{3}\right)-S_{6}\left(\boldsymbol{y}_{3},\,\boldsymbol{y}_{4},\,\boldsymbol{y}_{1},\,\boldsymbol{y}_{2},\,\boldsymbol{x}_{2},\,\boldsymbol{x}_{1}\right)S_{2}\left(\boldsymbol{x}_{4},\,\boldsymbol{x}_{3}\right)\right]\nonumber 
\end{align}
\begin{align}
 & T_{a}^{(4')}\left(\boldsymbol{x}_{1},\,\boldsymbol{x}_{2},\,\boldsymbol{x}_{3},\,\boldsymbol{x}_{4}\right)\left(T_{a}^{(4)}\left(\boldsymbol{y}_{1},\,\boldsymbol{y}_{2},\,\boldsymbol{y}_{3},\,\boldsymbol{y}_{4}\right)\right)^{*}=\nonumber \\
 & \frac{N_{c}^{2}}{4}\left[S_{8}\left(\boldsymbol{x}_{1},\,\boldsymbol{x}_{2},\,\boldsymbol{x}_{3},\,\boldsymbol{x}_{4},\,\boldsymbol{y}_{2},\,\boldsymbol{y}_{1},\,\boldsymbol{y}_{4},\,\boldsymbol{y}_{3}\right)+S_{4}\left(\boldsymbol{x}_{1},\,\boldsymbol{x}_{2},\,\boldsymbol{y}_{2},\,\boldsymbol{y}_{1}\right)S_{2}\left(\boldsymbol{x}_{4},\,\boldsymbol{x}_{3}\right)S_{2}\left(\boldsymbol{y}_{4},\,\boldsymbol{y}_{3}\right)-\right.\\
 & -\left.S_{6}\left(\boldsymbol{x}_{1},\,\boldsymbol{x}_{2},\,\boldsymbol{x}_{3},\,\boldsymbol{x}_{4},\,\boldsymbol{y}_{2},\,\boldsymbol{y}_{1}\right)S_{2}\left(\boldsymbol{y}_{4},\,\boldsymbol{y}_{3}\right)-S_{6}\left(\boldsymbol{y}_{3},\,\boldsymbol{y}_{4},\,\boldsymbol{y}_{1},\,\boldsymbol{y}_{2},\,\boldsymbol{x}_{2},\,\boldsymbol{x}_{1}\right)S_{2}\left(\boldsymbol{x}_{4},\,\boldsymbol{x}_{3}\right)\right]\nonumber 
\end{align}
\begin{align}
 & T_{a}^{(4)}\left(\boldsymbol{x}_{1},\,\boldsymbol{x}_{2},\,\boldsymbol{x}_{3},\,\boldsymbol{x}_{4}\right)\left(T_{a}^{(4')}\left(\boldsymbol{y}_{1},\,\boldsymbol{y}_{2},\,\boldsymbol{y}_{3},\,\boldsymbol{y}_{4}\right)\right)^{*}=\nonumber \\
 & \frac{N_{c}^{2}}{4}\left[S_{8}\left(\boldsymbol{x}_{3},\,\boldsymbol{x}_{4},\,\boldsymbol{x}_{1},\,\boldsymbol{x}_{2},\,\boldsymbol{y}_{4},\,\boldsymbol{y}_{3},\,\boldsymbol{y}_{2},\,\boldsymbol{y}_{1}\right)+S_{4}\left(\boldsymbol{x}_{1},\,\boldsymbol{x}_{2},\,\boldsymbol{y}_{2},\,\boldsymbol{y}_{1}\right)S_{2}\left(\boldsymbol{x}_{4},\,\boldsymbol{x}_{3}\right)S_{2}\left(\boldsymbol{y}_{4},\,\boldsymbol{y}_{3}\right)-\right.\\
 & -\left.S_{6}\left(\boldsymbol{x}_{3},\,\boldsymbol{x}_{4},\,\boldsymbol{x}_{1},\,\boldsymbol{x}_{2},\,\boldsymbol{y}_{2},\,\boldsymbol{y}_{1}\right)S_{2}\left(\boldsymbol{y}_{4},\,\boldsymbol{y}_{3}\right)-S_{6}\left(\boldsymbol{y}_{4},\,\boldsymbol{y}_{3},\,\boldsymbol{y}_{2},\,\boldsymbol{y}_{1},\,\boldsymbol{x}_{2},\,\boldsymbol{x}_{1}\right)S_{2}\left(\boldsymbol{x}_{4},\,\boldsymbol{x}_{3}\right)\right]\nonumber 
\end{align}
\begin{align}
 & T_{a}^{(4')}\left(\boldsymbol{x}_{1},\,\boldsymbol{x}_{2},\,\boldsymbol{x}_{3},\,\boldsymbol{x}_{4}\right)\left(T_{a}^{(4')}\left(\boldsymbol{y}_{1},\,\boldsymbol{y}_{2},\,\boldsymbol{y}_{3},\,\boldsymbol{y}_{4}\right)\right)^{*}=\nonumber \\
 & \frac{N_{c}^{2}}{4}\left[S_{8}\left(\boldsymbol{x}_{1},\,\boldsymbol{x}_{2},\,\boldsymbol{x}_{3},\,\boldsymbol{x}_{4},\,\boldsymbol{y}_{4},\,\boldsymbol{y}_{3},\,\boldsymbol{y}_{2},\,\boldsymbol{y}_{1}\right)+S_{4}\left(\boldsymbol{x}_{1},\,\boldsymbol{x}_{2},\,\boldsymbol{y}_{2},\,\boldsymbol{y}_{1}\right)S_{2}\left(\boldsymbol{x}_{4},\,\boldsymbol{x}_{3}\right)S_{2}\left(\boldsymbol{y}_{4},\,\boldsymbol{y}_{3}\right)-\right.\\
 & -\left.S_{6}\left(\boldsymbol{x}_{1},\,\boldsymbol{x}_{2},\,\boldsymbol{x}_{3},\,\boldsymbol{x}_{4}\,\boldsymbol{y}_{2},\,\boldsymbol{y}_{1}\right)S_{2}\left(\boldsymbol{y}_{4},\,\boldsymbol{y}_{3}\right)-S_{6}\left(\boldsymbol{y}_{4},\,\boldsymbol{y}_{3},\,\boldsymbol{y}_{2},\,\boldsymbol{y}_{1},\,\boldsymbol{x}_{2},\,\boldsymbol{x}_{1}\right)S_{2}\left(\boldsymbol{x}_{4},\,\boldsymbol{x}_{3}\right)\right]\nonumber 
\end{align}
We can see that the SPS cross-section~(\ref{eq:ASq}) is expressed
in terms of dipoles, quadrupoles, sextupoles and octupole scattering
amplitudes. 

\subsubsection{Fragmentation of heavy quarks into heavy quarkonia}

\label{subsec:Fragmentation}
\begin{figure}
\includegraphics[width=5cm]{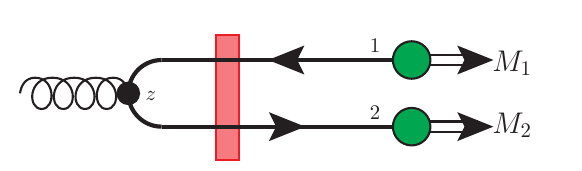}\includegraphics[width=5cm]{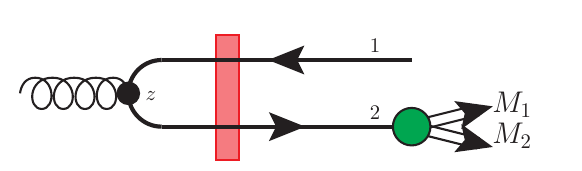}

\caption{\protect\label{fig:CGCBasic-1}The diagrams which describe the heavy
quarkonia pair production via single- and dihadron fragmentation in
the leading order in $\alpha_{s}$. The red block represents the interaction
with the target (shockwave).}
\end{figure}
The fragmentation mechanism formally is also induced by a single-gluon
scattering, with subsequent fragmentation of the produced heavy quarks
into quarkonia states, as shown schematically in the Figure~~\ref{fig:CGCBasic-1}.
A naive evaluation of the impact factor suggests that this mechanism
should be dominant in the leading order in $\alpha_{s}$, exceeding
the contribution found in the previous section. However, as was discussed
in~\cite{Ma:2013yla}, the fragmentation functions in the heavy
quark mass limit introduce additional suppression by factor $\sim\alpha_{s}^{2}\left(m_{Q}\right)$,
and for this reason the cross-section of this mechanism is suppressed
by factor $\sim\alpha_{s}^{2}\left(m_{Q}\right)$ compared to other
mechanisms. Nevertheless, for the sake of completeness, we will provide
in this subsection the analytic results for the cross-section of the
process. Both for the single-hadron and dihadron fragmentation shown
schematically in the Figure~~\ref{fig:CGCBasic-1} the cross-section
can be represented as a convolution of the heavy quark pair production
cross-section, and the dihadron or a product of two single-hadron
fragmentation functions, namely
\begin{align}
\frac{d\sigma_{T}\left(p+p\to M_{1}+M_{2}+X\right)}{dy_{1}d^{2}\boldsymbol{p}_{1}^{\perp}dy_{2}d^{2}\boldsymbol{p}_{2}^{\perp}} & =\int_{0}^{1}\frac{dz_{1}}{z_{1}^{2}}\int_{0}^{1}\frac{dz_{2}}{z_{2}^{2}}\left[D_{f}\left(z_{1}\right)D_{f}\left(z_{2}\right)+D_{f}\left(z_{1},\,z_{2}\right)\right]\frac{d\sigma_{T}\left(p+p\to Q_{1}+Q_{2}+X\right)}{d\mathfrak{y}_{1}d^{2}\boldsymbol{\ell}_{1,\perp}d\mathfrak{y}_{2}\boldsymbol{\ell}_{2,\perp}}\label{eq:sigma}
\end{align}
where $D_{f}(z)$ and $D_{f}\left(z_{1},\,z_{2}\right)$ are the corresponding
single- and dihadron fragmentation functions, and $z_{i}$ are the
fractions of the quark momentum carried by the quarkonium. At present
the dihadron fragmentation function $D_{f}\left(z_{1},\,z_{2}\right)$
is not known. In the heavy quark mass limit we may expect that for
the charmonia-bottomonia pairs the dihadron fragmentation function
are very small~\footnote{If we assume that a perturbative treatment is applicable for evaluation
of the dihadron fragmentation function, it is possible to show that
in the heavy quark mass limit the dihadron fragmentation function
is suppressed by $\sim m_{Q}^{-4}$ compared to the product of two
single-hadron fragmentation functions due to additional propagators.}, so for the sake of simplicity we will disregard its contribution
and focus on the first term in the brackets of~(\ref{eq:sigma}).
The tilde in the right-hand side of~(\ref{eq:sigma}) is used to
emphasize that the corresponding rapidities and momenta of heavy quarks
should be adjusted in order to take into account that the observed
heavy quark has only fraction of the heavy quark momentum. For independent
fragmentation from two heavy quarks this adjustment is given by 
\begin{equation}
\mathfrak{y}_{i}=y_{i}+\ln\left(1/z_{i}\right),\quad\boldsymbol{\ell}_{i,\perp}=\boldsymbol{p}_{i}^{\perp}/z_{i}.
\end{equation}
We need to mention that this adjustment implies that the integrals
over $z_{1},z_{2}$ in~(\ref{eq:sigma}) should be regularized (cut)
at the lower limit, which may be fixed from the condition that the
sum of energies of two heavy quarks should no exceed the energy of
the incoming gluon. However, in the high energy kinematics we expect
that the energy of the incoming gluon is very large, so this correction
can be disregarded.

The cross-section of the $pp\to\bar{Q}Q\,X$ subprocess has been found
in the CGC framework~\cite{Fujii:2020bkl} and is given by
\begin{align}
\frac{d\sigma\left(p+p\to Q_{1}+Q_{2}+X\right)}{d\mathfrak{y}_{1}d^{2}\boldsymbol{\ell}_{1,\perp}d\mathfrak{y}_{2}\boldsymbol{\ell}_{2,\perp}} & =\frac{x_{1}g\left(x_{1}\right)}{\left(2\pi\right)^{4}}\int d^{8}\boldsymbol{X}e^{-i\left(\boldsymbol{\ell}_{1\perp}+\boldsymbol{\ell}_{2\perp}\right)\cdot\left(\boldsymbol{b}_{12}-\boldsymbol{b}_{12}'\right)}e^{-i\left(\bar{\alpha}\boldsymbol{\ell}_{1\perp}-\alpha\boldsymbol{\ell}_{2\perp}\right)\cdot\left(\boldsymbol{r}_{12}-\boldsymbol{r}_{12}'\right)}\times\label{eq:sixfold}\\
 & \times\,\alpha(1-\alpha)\,\sum_{\lambda}R_{\lambda}\left(\alpha,\,\boldsymbol{r}_{12},\,\boldsymbol{r}_{12}',\,Q^{2}\right)\,\Xi_{{\rm LO}}\left(\boldsymbol{x}_{1},\,\boldsymbol{x}_{2},\,\boldsymbol{x}_{3},\,\boldsymbol{x}_{4}\right),\nonumber 
\end{align}
where $d^{8}\boldsymbol{X}$ implies integration over the transverse
coordinates of the heavy quarks in the amplitude and its conjugate
(we use notations $\boldsymbol{x}_{1},\boldsymbol{x}_{2}$ for the
coordinates in the amplitude, and $\boldsymbol{x}_{1}',\boldsymbol{x}_{2}'$
for its conjugate). We also defined the relative distances and pairwise
impact parameters as
\begin{align}
\boldsymbol{r}_{12} & =\boldsymbol{x}_{1}-\boldsymbol{x}_{2},\quad\boldsymbol{r}_{12}'=\boldsymbol{x}_{1}'-\boldsymbol{x}_{2}',\quad\boldsymbol{b}_{12}=\alpha\boldsymbol{x}_{1}+\bar{\alpha}\boldsymbol{x}_{2},\quad\boldsymbol{b}_{12}'=\alpha\boldsymbol{x}_{1}'+\bar{\alpha}\boldsymbol{x}_{2}'.
\end{align}
where the variable $\alpha$ is the light-cone fraction of the gluon
momentum carried by the quark, 
\begin{equation}
\alpha=\frac{m_{1,\perp}e^{\mathfrak{y}_{1}}}{m_{1,\perp}e^{\mathfrak{y}_{1}}+m_{2,\perp}e^{\mathfrak{y}_{2}}},\qquad{\rm where}\qquad m_{a,\perp}=\sqrt{m_{b}^{2}+\left(\boldsymbol{\ell}_{a,\perp}\right)^{2}}.
\end{equation}
The impact factor $R_{\lambda}$ can be evaluated using the CGC rules
from the Table~\ref{tab:FR} and the general procedure outlined at
the end of the Section~\ref{subsec:Derivation}. The final result
of this procedure yields~\cite{Dosch:1996ss,Bjorken:1970ah}

\begin{align}
R_{T}\left(\alpha,\,\boldsymbol{r}_{12},\,\boldsymbol{r}_{12}',\,Q^{2}\right) & =\frac{\alpha_{s}}{4\pi^{2}}\left\{ \epsilon_{f}^{2}\,K_{1}\left(\epsilon_{f}r_{12}\right)K_{1}\left(\epsilon_{f}r_{12}'\right)\left[\alpha^{2}+(1-\alpha)^{2}\right]\hat{\boldsymbol{r}}_{12}\cdot\hat{\boldsymbol{r}}_{12}'\right.\left.+m_{f}^{2}K_{0}\left(\epsilon_{f}r_{12}\right)K_{0}\left(\epsilon_{f}r_{12}'\right)\right\} \\
R_{L}\left(\alpha,\,\boldsymbol{r}_{12},\,\boldsymbol{r}_{12}',\,Q^{2}\right) & =\frac{\alpha_{s}}{4\pi^{2}}\,\left\{ 4Q^{2}\alpha^{2}(1-\alpha)^{2}K_{0}\left(\epsilon_{f}r_{12}\right)K_{0}\left(\epsilon_{f}r_{12}'\right)\right\} .
\end{align}
The interaction with the target in the leading does not depend on
the helicity of the incoming gluon and in the large-$N_{c}$ limit
reduces to
\begin{align}
\Xi_{{\rm LO}}= & \frac{1}{N_{c}^{2}-1}\left[\frac{}{}-S_{4}\left(Y,\boldsymbol{x}_{1},\,\boldsymbol{x}_{2},\,\boldsymbol{x}_{2}',\,\boldsymbol{x}_{1}'\right)+N_{c}^{2}S_{2}\left(\boldsymbol{x}_{2}',\boldsymbol{x}_{2}\right)S_{2}\left(\boldsymbol{x}_{1},\,\boldsymbol{x}_{1}'\right)\right.-N_{c}^{2}S_{2}\left(\boldsymbol{x}_{1},\,\boldsymbol{x}_{12}'\right)S_{2}\left(\boldsymbol{x}_{12}',\,\boldsymbol{x}_{2}\right)\\
 & -N_{c}^{2}S_{2}\left(\boldsymbol{x}_{1}',\,\boldsymbol{x}_{12}\right)S_{2}\left(\boldsymbol{x}_{12},\,\boldsymbol{x}_{2}'\right)+S_{2}\left(\boldsymbol{x}_{1}',\,\boldsymbol{x}_{2}'\right)+S_{2}\left(\boldsymbol{x}_{1},\,\boldsymbol{x}_{2}\right)\left.+N_{c}^{2}S_{2}\left(\boldsymbol{x}_{12},\,\boldsymbol{x}_{12}'\right)S_{2}\left(\boldsymbol{x}_{12}',\,\boldsymbol{x}_{12}\right)-1\frac{}{}\right].\nonumber 
\end{align}
 In view of the aforementioned formal suppression of this mechanism
and the expected numerical smallness, as well as lack of reliable
parametrizations of the dihadron fragmentation functions, in what
follows we will disregard the contributions of this mechanism. 

\subsection{Charmonia-bottomonia production via Double Parton Scattering}

\label{subsec:DPS}
\begin{figure}
\includegraphics[width=3.8cm]{Figures/FIG_1B_06a_Rename_g_DPS}\includegraphics[width=3.8cm]{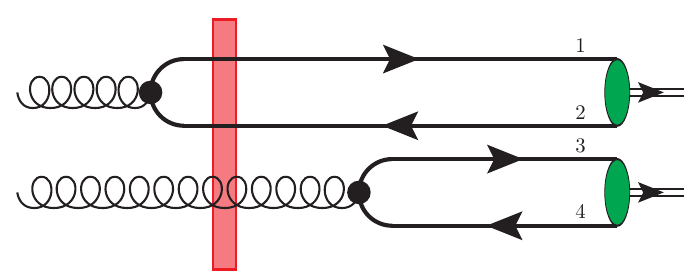}\includegraphics[width=3.8cm]{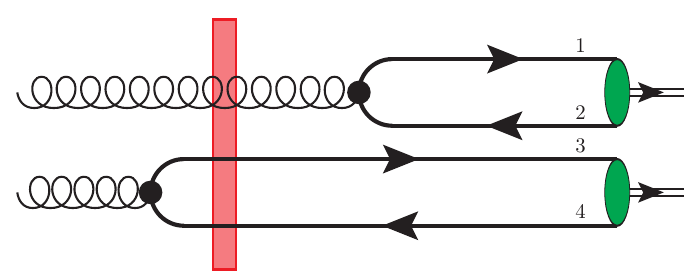}\includegraphics[width=3.8cm]{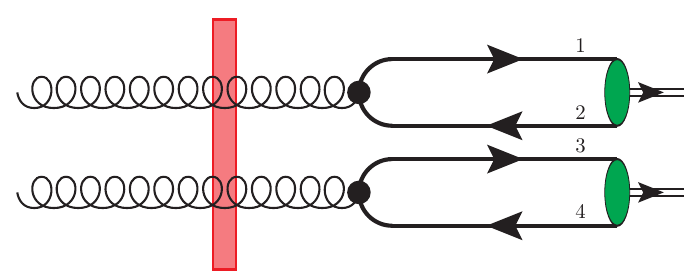}

\caption{\protect\label{fig:CGCBasic-3}The diagrams which describe the inclusive
charmonia-bottomonia pair production from a pair of the primordial
gluons (Double Parton Scattering mechanism). The diagrams 2-4 include
production of one of the heavy $\bar{Q}Q$ pairs in color octet state
and thus can be disregarded due to smallness of color octet LDME.
For the quarkonia pairs with the same hidden flavor, this mechanism
should be supplemented with additional contribution in which the heavy
quarks $(1,3)$ are permuted. The red block represents the interaction
with the target (shockwave).}
\end{figure}
The production of the charmonia-bottomonia pairs via the double parton
scattering proceeds via mechanism that can be schematically represented
by the diagrams shown in the Figure~\ref{fig:CGCBasic-3}. As explained
in the previous section, due to smallness of the color octet LDMEs
it is possible to disregard completely the contributions of the diagrams
2-4 and focus only on the first diagram~\footnote{A full expression with account of color octet contributions may be
found in~\cite{Kang:2013hta}}. The evaluation of the cross-section largely repeats a similar evaluation
of the inclusive single quarkonia production from~\cite{Kang:2013hta}.
The final result for the cross-section may be represented as 
\begin{align}
\frac{d\sigma\left(p+p\to M_{1}+M_{2}+X\right)}{dy_{1}d^{2}\boldsymbol{p}_{1}^{\perp}dy_{2}d^{2}\boldsymbol{p}_{2}^{\perp}} & =\frac{xg_{f}\left(x_{1},\,x_{2}\right)}{4}\,\,\int\prod_{k=1}^{4}d^{2}\boldsymbol{x}_{k}d^{k}\boldsymbol{y}_{k}e^{-i\left(\boldsymbol{p}_{1}-\boldsymbol{\mathfrak{q}}_{1,\perp}\right)\cdot\left(\boldsymbol{x}_{1}+\boldsymbol{x}_{2}+\boldsymbol{y}_{1}+\boldsymbol{y}_{2}\right)}e^{-i\left(\boldsymbol{p}_{2}-\boldsymbol{\mathfrak{q}}_{2,\perp}\right)\cdot\left(\boldsymbol{x}_{3}+\boldsymbol{x}_{4}+\boldsymbol{y}_{3}+\boldsymbol{y}_{4}\right)}\times\label{eq:XSec}\\
 & \times R_{M_{1}}\left(\boldsymbol{x}_{1},\boldsymbol{x}_{2},\boldsymbol{y}_{1},\boldsymbol{y}_{2}\right)R_{M_{2}}\left(\boldsymbol{x}_{3},\boldsymbol{x}_{4},\boldsymbol{y}_{3},\boldsymbol{y}_{4}\right)\times\nonumber \\
 & \times\left[S_{4}\left(\boldsymbol{x}_{1},\boldsymbol{x}_{2},\boldsymbol{y}_{1},\boldsymbol{y}_{2}\right)-S_{2}\left(\boldsymbol{x}_{1},\boldsymbol{x}_{2}\right)S_{2}\left(\boldsymbol{y}_{1},\boldsymbol{y}_{2}\right)\right]\left[S_{4}\left(\boldsymbol{x}_{3},\boldsymbol{x}_{4},\boldsymbol{y}_{3},\boldsymbol{y}_{4}\right)-S_{2}\left(\boldsymbol{x}_{3},\boldsymbol{x}_{4}\right)S_{2}\left(\boldsymbol{y}_{3},\boldsymbol{y}_{4}\right)\right]\nonumber 
\end{align}
where $g_{f}\left(x_{1},\,x_{2}\right)$ is the so-called double parton
distribution function (DPDF) of the gluon, the impact factors $R_{M_{1}}$
and $R_{M_{2}}$ are the impact factors of the single-quarkonium production
found in~\cite{Kang:2013hta}. After adjustment of notations and
omission of the relative motion of the quarks in the quarkonium, the
impact factors for the $S$-wave quarkonia are given explicitly by
\begin{align}
R_{M_{a}}\left(\boldsymbol{x}_{1},\boldsymbol{x}_{2},\boldsymbol{y}_{1},\boldsymbol{y}_{2}\right) & =\mathcal{M}_{M_{a}}^{*}\left(\boldsymbol{y}_{1},\boldsymbol{y}_{2}\right)\mathcal{M}_{M_{a}}\left(\boldsymbol{x}_{1},\boldsymbol{x}_{2}\right),\qquad a=1,2\\
\mathcal{M}_{M_{a}}\left(\boldsymbol{x}_{1},\boldsymbol{x}_{2}\right) & =\left(ig\right)\int\frac{dk^{-}d^{2}\boldsymbol{k}}{\left(2\pi\right)^{3}}e^{i\left(\boldsymbol{k}-\frac{\boldsymbol{p}_{a}^{\perp}-\boldsymbol{\mathfrak{q}}_{a,\perp}}{2}\right)\cdot\left(\boldsymbol{x}_{1}-\boldsymbol{x}_{2}\right)}\,{\rm Tr}\left[\Pi_{M_{a}}^{S,S_{z}}\gamma^{+}S\left(\frac{p_{a}}{2}-k\right)\hat{\varepsilon}(\mathfrak{q}_{a})S\left(\frac{p_{a}}{2}-k-\mathfrak{q}_{a}\right)\gamma^{+}\right]=\\
 & =-\left(ig\right)\int\frac{d^{2}\boldsymbol{k}}{\left(2\pi\right)^{2}}e^{-i\boldsymbol{k}\cdot\left(\boldsymbol{x}_{1}-\boldsymbol{x}_{2}\right)/2}\,\frac{{\rm Tr}\left[\Pi_{M_{a}}^{S,S_{z}}\gamma^{+}\left(\frac{\hat{p}_{a}}{2}-\hat{k}+m_{a}\right)\hat{\varepsilon}(\mathfrak{q}_{a})\left(\frac{\hat{p}_{a}}{2}-\hat{k}-\hat{\mathfrak{q}}_{a}+m_{a}\right)\gamma^{+}\right]}{p_{a}^{+}\left[\left(\frac{\boldsymbol{p}_{a}^{\perp}}{2}-\boldsymbol{k}\right)^{2}+m_{a}^{2}\right]+p_{a}^{+}\left[\left(\frac{\boldsymbol{p}_{a}^{\perp}}{2}-\boldsymbol{k}-\boldsymbol{\mathfrak{q}}_{a,\perp}\right)^{2}+m_{a}^{2}\right]}\nonumber 
\end{align}
where $\varepsilon(\mathfrak{q}_{a})$ is the polarization vector
of the incoming gluon, the 4-vector $k^{\mu}$ is defined as change
of the quark momentum after interaction with the shock wave, and at
the last step we applied the Cauchy theorem for the integration over
$k^{-}$. We also took into account that in eikonal approximation
$\mathfrak{q}_{a}^{+}=p_{a}^{+}$ since the interaction with the shockwave
does not change the plus-component of the momentum. Using the procedure
described at the end of Section~\ref{subsec:Derivation} and introducing
the helicity projectors $\bar{\Pi}_{M_{a}}^{h_{1},h_{2}}$ defined
in~(\ref{eq:P1}), we may obtain

\begin{align}
\mathcal{M}_{M_{a}}\left(\boldsymbol{x}_{1},\boldsymbol{x}_{2}\right) & =\bar{\Pi}_{M_{1}}^{h_{1},h_{2}}\int\frac{d^{2}\boldsymbol{k}}{\left(2\pi\right)^{2}}\,\frac{\left(-ig\right){\rm Tr}\left[\bar{u}_{h_{1}}\left(\frac{\hat{p}_{a}}{2}-\hat{k}\right)\hat{\varepsilon}(\mathfrak{q}_{a})v_{h_{2}}\left(\frac{p_{a}}{2}-k-\mathfrak{q}_{a}\right)\right]e^{i\left(\boldsymbol{k}-\frac{\boldsymbol{p}_{a}^{\perp}-\boldsymbol{\mathfrak{q}}_{a,\perp}}{2}\right)\cdot\left(\boldsymbol{x}_{1}-\boldsymbol{x}_{2}\right)}}{p_{a}^{+}\left[\left(\frac{\boldsymbol{p}_{a}^{\perp}}{2}-\boldsymbol{k}\right)^{2}+m_{a}^{2}\right]+p_{a}^{+}\left[\left(\frac{\boldsymbol{p}_{a}^{\perp}}{2}-\boldsymbol{k}-\boldsymbol{\mathfrak{q}}_{a,\perp}\right)^{2}+m_{a}^{2}\right]}.
\end{align}
The integral over $\boldsymbol{k}_{\perp}$ yields well-known result
for the wave function of $\bar{Q}Q$ in the photon evaluated in~\cite{Bjorken:1970ah,Dosch:1996ss},
namely
\begin{align}
 & \int\frac{d^{2}\boldsymbol{k}}{\left(2\pi\right)^{2}}\,\frac{\left(-ig\right){\rm Tr}\left[\bar{u}_{h_{1}}\left(\frac{p_{a}}{2}-k\right)\hat{\varepsilon}(\mathfrak{q}_{a})v_{h_{2}}\left(\mathfrak{q}_{a}+k-\frac{p_{a}}{2}\right)\right]e^{i\left(\boldsymbol{k}-\frac{\boldsymbol{p}_{a}^{\perp}-\boldsymbol{\mathfrak{q}}_{a,\perp}}{2}\right)\cdot\left(\boldsymbol{x}_{1}-\boldsymbol{x}_{2}\right)}}{p_{a}^{+}\left[\left(\frac{\boldsymbol{p}_{a}^{\perp}}{2}-\boldsymbol{k}\right)^{2}+m^{2}\right]+p_{a}^{+}\left[\left(\frac{\boldsymbol{p}_{a}^{\perp}}{2}-\boldsymbol{k}-\boldsymbol{\mathfrak{q}}_{a,\perp}\right)^{2}+m^{2}\right]}=\\
 & \sim\frac{-ig}{(2\pi)}\left[\left(\delta_{\lambda_{a},h_{1}}-\delta_{\lambda_{a},-h_{1}}\right)\delta_{h_{1},-h_{2}}i\boldsymbol{\varepsilon}_{\lambda_{a}}\cdot\nabla-m\sqrt{2}\,{\rm sign}(h_{1})\delta_{\lambda_{a},h_{1}}\delta_{h_{1},h_{2}}\right]K_{0}\left(\bar{m}_{a}\,r_{12}\right),\quad r_{12}=\left|\boldsymbol{x}_{1}-\boldsymbol{x}_{2}\right|\nonumber 
\end{align}
where $\lambda_{a}$ is the helicity of the incoming gluon and $\bar{m}_{a}=\sqrt{m_{a}^{2}+\mathfrak{\boldsymbol{q}}_{a}^{2}/4}$.
The interaction of the partonic ensemble with the target (shockwave)
is encoded in the correlator of Wilson lines which appears in the
last line of~(\ref{eq:XSec}). In the large-$N_{c}$ limit and in
heavy nuclei it is possible to factorize it and rewrite as
\begin{align}
 & \left\langle \left[S_{4}\left(\boldsymbol{x}_{1},\boldsymbol{x}_{2},\boldsymbol{y}_{1},\boldsymbol{y}_{2}\right)-S_{2}\left(\boldsymbol{x}_{1},\boldsymbol{x}_{2}\right)S_{2}\left(\boldsymbol{y}_{1},\boldsymbol{y}_{2}\right)\right]\left[S_{4}\left(\boldsymbol{x}_{3},\boldsymbol{x}_{4},\boldsymbol{y}_{3},\boldsymbol{y}_{4}\right)-S_{2}\left(\boldsymbol{x}_{3},\boldsymbol{x}_{4}\right)S_{2}\left(\boldsymbol{y}_{3},\boldsymbol{y}_{4}\right)\right]\right\rangle =\label{eq:Second}\\
 & =\kappa\left\langle \left[S_{4}\left(\boldsymbol{x}_{1},\boldsymbol{x}_{2},\boldsymbol{y}_{1},\boldsymbol{y}_{2}\right)-S_{2}\left(\boldsymbol{x}_{1},\boldsymbol{x}_{2}\right)S_{2}\left(\boldsymbol{y}_{1},\boldsymbol{y}_{2}\right)\right]\right\rangle \left\langle \left[S_{4}\left(\boldsymbol{x}_{3},\boldsymbol{x}_{4},\boldsymbol{y}_{3},\boldsymbol{y}_{4}\right)-S_{2}\left(\boldsymbol{x}_{3},\boldsymbol{x}_{4}\right)S_{2}\left(\boldsymbol{y}_{3},\boldsymbol{y}_{4}\right)\right]\right\rangle .\nonumber 
\end{align}
where $\kappa$ is some numerical parameter, and angular brackets
imply averaging defined in~(\ref{eq:Ave}). We may observe that the
second line of~(\ref{eq:Second}) essentially factorizes onto a product
of the structures $\left\langle S_{4}-S_{2}\otimes S_{2}\right\rangle $
which appear in the single-quarkonium production~\cite{Kang:2013hta}.
Furthermore, as was discussed in~\cite{Golec-Biernat:2015az}, at
high energies it is possible to disregard small correlations of the
gluons in the projectile and approximate the gluonic DPDF in the first
line of~(\ref{eq:XSec}) as
\begin{equation}
xg_{f}\left(x_{1},\,x_{2}\right)\approx xg_{f}\left(x_{1}\right)\,\,xg_{f}\left(x_{2}\right).
\end{equation}
In this approximation we may recover the so-called pocket formula
\begin{equation}
\frac{d\sigma\left(p+p\to M_{1}+M_{2}+X\right)}{dy_{1}d^{2}\boldsymbol{p}_{1}^{\perp}dy_{2}d^{2}\boldsymbol{p}_{2}^{\perp}}\approx\kappa\frac{d\sigma\left(p+p\to M_{1}+X\right)}{dy_{1}d^{2}\boldsymbol{p}_{1}^{\perp}}\frac{d\sigma\left(p+p\to M_{2}+X\right)}{dy_{2}d^{2}\boldsymbol{p}_{2}^{\perp}}
\end{equation}
where the parameter $\kappa$ is usually replaced with the so-called
(inverse) effective cross-section, 
\begin{equation}
\kappa=\sigma_{{\rm eff}}^{-1},
\end{equation}
which was introduced in the context of double parton scattering in
the$k_{T}$ factorization framework~(see the review \cite{Lansberg:2019adr}).
These findings demonstrate that description of the double parton scattering
in CGC picture largely agrees with results of other frameworks. 

\section{Numerical estimates for the cross-section}

\label{sec:Numer}\label{sec:Parametrization-S4}

\subsection{Results for cross-sections}

For the sake of definiteness, we'll focus on the production of the
$1S$ quarkonia production, which are expected to have the largest
cross-section among all the charmonia-bottomonia pairs. We disregard
the color octet channels in view of the smallness of color octet LDMEs
(see however discussion in~\cite{Butenschoen:2014dra,Feng:2015cba,Baranov:2015laa,Baranov:2016clx}
about their significant uncertainty). For the dominant color singlet
LDMEs we will use the values~\cite{Feng:2015wka,Abdulov:2020nxh,Baranov:2016clx}
\begin{align}
\left\langle \mathcal{O}^{J/\psi}\left(^{3}S_{1}^{[1]}\right)\right\rangle  & \approx1.16\,{\rm GeV}^{3},\qquad\left\langle \mathcal{O}^{\Upsilon(1S)}\left(^{3}S_{1}^{[1]}\right)\right\rangle \approx8.39\,{\rm GeV}^{3}.\label{eq:CO1-2}
\end{align}
 In agreement with NRQCD and potential models~\cite{Eichten:1978tg,Eichten:1979ms,Quigg:1977dd,Martin:1980jx},
these LDMEs approximately scale with heavy quark mass as $\sim\left(\alpha_{s}\left(m_{Q}\right)m_{Q}\right)^{3}$.
As we demonstrated in the previous sections, the cross-section of
the charmonia-bottomonia production depends on a number of multipole
correlators $S_{2n}\left(\boldsymbol{x}_{1},...,\boldsymbol{x}_{2n}\right)$
defined in~\ref{eq:MultipoleDfinition}. For the dipole amplitude
($n=1$), we will use the phenomenological impact parameter $b$-dependent
``bCGC'' parametrization with the set parameters from~\cite{RESH}
corresponding to a charm mass $m_{c}\approx1.4$~GeV, 
\begin{align}
N\left(x,\,\boldsymbol{r},\,\boldsymbol{b}\right) & =\left\{ \begin{array}{cc}
N_{0}\,\left(\frac{r\,Q_{s}(x)}{2}\right)^{2\gamma_{{\rm eff}}(r)}, & r\,\le\frac{2}{Q_{s}(x)}\\
1-\exp\left(-\mathcal{A}\,\ln\left(\mathcal{B}r\,Q_{s}\right)\right), & r\,>\frac{2}{Q_{s}(x)}
\end{array}\right.~,\label{eq:CGCDipoleParametrization}\\
 & \mathcal{A}=-\frac{N_{0}^{2}\gamma_{s}^{2}}{\left(1-N_{0}\right)^{2}\ln\left(1-N_{0}\right)},\quad\mathcal{B}=\frac{1}{2}\left(1-N_{0}\right)^{-\frac{1-N_{0}}{N_{0}\gamma_{s}}},\\
 & Q_{s}(x,\,\boldsymbol{b})=\left(\frac{x_{0}}{x}\right)^{\lambda/2}T_{G}(b),\,\,\gamma_{{\rm eff}}(r)=\gamma_{s}+\frac{1}{\kappa\lambda Y}\ln\left(\frac{2}{r\,Q_{s}(x)}\right),\,\,T_{G}(b)=\exp\left(-\frac{b^{2}}{4\gamma_{s}B_{{\rm CGC}}}\right)\label{eq:gamma_eff}\\
\gamma_{s} & =0.6492,\quad\lambda=0.2023,\quad x_{0}=6.9\times10^{-4},\quad B_{{\rm CGC}}=5.5\,{\rm GeV}^{-2}\approx\left(0.463\,{\rm fm}\right)^{2}.
\end{align}

The higher order amplitudes in general are independent nonperturbative
objects, which can be related to a dipole amplitude under additional
assumptions. In the McLerran-Venugopalan model, the weighting functional
$W[\rho]$ has a Gaussian form, and for this reason all $2n$-leg
correlators $S_{2n}\left(\boldsymbol{x}_{1},...,\boldsymbol{x}_{2n}\right)$
reduce to pairwise products of dipole amplitudes $S_{2}$, summed
over different permutations of $\boldsymbol{x}_{1},....\boldsymbol{x}_{2n}$,
namely
\begin{align}
\left\langle U^{\dagger}\left(\boldsymbol{y}_{1}\right)^{a_{1}b_{1}}U\left(\boldsymbol{y}_{2}\right)^{a_{2}b_{2}}...U^{\dagger}\left(\boldsymbol{y}_{2n-1}\right)^{a_{2n-1}b_{2n-1}}U\left(\boldsymbol{y}_{2n}\right)^{a_{2n}b_{2n}}\right\rangle  & \approx\sum_{\sigma\in\Pi(\chi)}\prod_{\{\alpha,\beta\}\in\sigma}\left\langle U^{\dagger}\left(\boldsymbol{y}_{\alpha}\right)^{a_{\alpha}b_{\alpha}}U\left(\boldsymbol{y}_{\beta}\right)^{a_{\beta}b_{\beta}}\right\rangle ,\label{eq:Wick-1}
\end{align}
where $\chi=\{1,2,...,2n\}$ and $\Pi(\chi)$ is the set of partitions
of $\chi$ with disjoint pairs. The result~(\ref{eq:Wick-1}) essentially
allows to use the Wick's theorem for evaluation of the diagrams with
multipole contributions.  The expression~(\ref{eq:Wick-1}) sometimes
is also justified using the Kovner's \textit{area enhancement argument}
introduced in~\cite{Agostini:2021xc,Kovner:2017ssr,Kovner:2018vec}:
the color group averaging suppresses all configurations except those
in which a small group of partons have a net zero color charge and
are tightly clustered, with relative distances constrained by $\left|\boldsymbol{x}_{i}-\boldsymbol{x}_{j}\right|\lesssim Q_{s}^{-1}$,
where $Q_{s}$ is the saturation scale; in the phase space integral
the largest phase space have color singlet quark-antiquark pairs.
The advantage of the area law argument is that it does not rely on
specific choice of the weighting functional $W[\rho]$. However, the
ansatz~(\ref{eq:Wick-1}) may be not compatible with JIMWLK evolution
equation: as was found in~\cite{Iancu:2011nj,Iancu:2011ns,Iancu:2012xa}
solving the JIMWLK equations for the quadrupoles in the limit of large
number of colors $N_{c}\gg1$, the corresponding solution can be described
in terms of dipole amplitudes, though the general expression is much
more complicated and does not reduce to~(\ref{eq:Wick-1}). Similar
solutions of the JIMWLK equations in the large-$N_{c}$ limit are
also known for the sextupoles and octupoles~\cite{Dominguez:2012ad,Shi:2017gcq},
and in general they also do not coincide with~(\ref{eq:Wick-1}).
The large-$N_{c}$ solution of JIMWLK equation coincides with~(\ref{eq:Wick-1})
for the configuration of small well-separated pairs of quark-antiquarks,
as expected from area enhancement argument. However, for charmonia-bottomonia
production we found that this argument does not work because the impact
factors of SPS production select configurations with approximately
the same distances between heavy partons. At the same time, using
the large-$N_{c}$ solutions directly is prohibitively expensive:
as could be seen from~(\ref{eq:ASq}), the evaluation of the cross-section
requires to take the integral in 16 dimensions, and the explicit expressions
for higher order multipoles from~\cite{Dominguez:2012ad,Shi:2017gcq}
include hundreds of terms with nontrivial behavior along various hypersurfaces
in 16-dimensional space. 

Fortunately, it is possible to simplify analysis in the so-called
dilute (weak scattering) limit~\cite{Iancu:2011nj,Iancu:2011ns},
which may be applicable at moderately high energies, as well as in
the heavy quark mass limit, when the mass $m_{Q}$ exceeds significantly
the saturation scale $Q_{s}$, and thus the distance between heavy
quarks is given by $\left|\boldsymbol{x}_{i}-\boldsymbol{x}_{j}\right|\sim m_{Q}^{-1}\ll Q_{s}^{-1}$.
In this regime we may approximate the Wilson lines~(\ref{eq:Wilson})
as 
\begin{equation}
U\left(\boldsymbol{x}\right)\approx1+ig\int dx^{-}A_{a}^{+}\left(x^{-},\,\boldsymbol{x}_{\perp}\right)t^{a}+\frac{(ig)^{2}}{2}P\int dx_{1}^{-}A_{a}^{+}\left(x_{1}^{-},\,\boldsymbol{x}_{\perp}\right)t^{a}\int dx_{1}^{-}A_{b}^{+}\left(x_{2}^{-},\,\boldsymbol{x}_{\perp}\right)t^{b}
\end{equation}
 and obtain for the dipole $S$-matrix element 
\begin{align}
S_{2}\left(Y,\,\boldsymbol{x}_{1},\,\boldsymbol{x}_{2}\right) & \equiv\frac{1}{N_{c}}\left\langle {\rm tr}\left(U\left(\boldsymbol{\boldsymbol{x}}_{1}\right)U^{\dagger}\left(\boldsymbol{x}_{2}\right)\right)\right\rangle _{Y}\approx1-\frac{\left|\alpha_{a}\left(\boldsymbol{x}_{1}\right)-\alpha_{a}\left(\boldsymbol{x}_{2}\right)\right|^{2}}{4N_{c}},\quad\alpha_{a}(\boldsymbol{x})=g\int dx^{-}A_{a}^{+}\left(x^{-},\,\boldsymbol{x}_{\perp}\right)t^{a}
\end{align}
so the forward dipole scattering amplitude can be approximated as
\begin{equation}
N\left(\boldsymbol{r}_{12},\,\boldsymbol{b}_{12}\right)\equiv1-S_{2}\left(\boldsymbol{x}_{1},\,\boldsymbol{x}_{2}\right)=\frac{\left|\alpha_{a}\left(\boldsymbol{x}_{1}\right)-\alpha_{a}\left(\boldsymbol{x}_{2}\right)\right|^{2}}{4N_{c}},
\end{equation}
where we defined the relative distance and the average center of mass
position using $\boldsymbol{r}_{ij}=\boldsymbol{x}_{i}-\boldsymbol{x}_{j},\quad\boldsymbol{b}_{ij}=\left(\alpha_{i}\boldsymbol{x}_{i}+\alpha_{j}\boldsymbol{x}_{j}\right)/(\alpha_{i}+\alpha_{j})$.
The expression for the quadrupole matrix element $S_{4}$ in this
limit is given by~\cite{Iancu:2011nj,Iancu:2011ns}

\begin{align}
S_{4} & \equiv\left\langle \frac{1}{N_{c}}{\rm tr}_{c}\left[U\left(\boldsymbol{x}_{1}\right)U^{\dagger}\left(\boldsymbol{x}_{2}\right)U\left(\boldsymbol{x}_{3}\right)U^{\dagger}\left(\boldsymbol{x}_{4}\right)\right]\right\rangle \approx1-\,N\left(x,\,\boldsymbol{r}_{12},\,\boldsymbol{b}_{12}\right)-\,N\left(x,\,\boldsymbol{r}_{34},\,\boldsymbol{b}_{34}\right)\label{eq:S4dilute-1}\\
 & -N\left(x,\,\boldsymbol{r}_{23},\,\boldsymbol{b}_{23}\right)-N\left(x,\,\boldsymbol{r}_{14},\,\boldsymbol{b}_{14}\right)+N\left(x,\,\boldsymbol{r}_{13},\,\boldsymbol{b}_{13}\right)+N\left(x,\,\boldsymbol{r}_{24},\,\boldsymbol{b}_{24}\right).\nonumber 
\end{align}
 The expressions for $T_{a}^{(4)},\,T_{a}^{(4')},T_{a}^{(8)}$ defined
in~(\ref{eq:T4}-\ref{eq:T8}) simplify in this limit as
\begin{align}
T_{a}^{(4)}\left(\boldsymbol{x}_{1},\,\boldsymbol{x}_{2},\,\boldsymbol{x}_{3},\,\boldsymbol{x}_{4}\right) & \approx T_{a}^{(4')}\left(\boldsymbol{x}_{1},\,\boldsymbol{x}_{2},\,\boldsymbol{x}_{3},\,\boldsymbol{x}_{4}\right)\approx\frac{i}{4}\left[\alpha_{a}\left(\boldsymbol{x}_{3}\right)-\alpha_{a}\left(\boldsymbol{x}_{4}\right)\right],
\end{align}
whereas the convolutions of two $T_{a}^{(4)}$ with different arguments
can be expressed in terms of the dipole forward scattering amplitude
as
\begin{equation}
\frac{8}{N_{c}}T_{a}^{(4)}\left(\boldsymbol{x}_{1},\,\boldsymbol{x}_{2},\,\boldsymbol{x}_{3},\,\boldsymbol{x}_{4}\right)T_{a}^{(4)}\left(\boldsymbol{x}_{1'},\,\boldsymbol{x}_{2'},\,\boldsymbol{x}_{3'},\,\boldsymbol{x}_{4'}\right)=N\left(\boldsymbol{r}_{3,3'},\,\boldsymbol{b}_{3,3'}\right)+N\left(\boldsymbol{r}_{4,4'},\,\boldsymbol{b}_{4,4'}\right)-N\left(\boldsymbol{r}_{3,4'},\,\boldsymbol{b}_{3,4'}\right)-N\left(\boldsymbol{r}_{4,3'},\,\boldsymbol{b}_{4,3'}\right)
\end{equation}
The combination of dipoles and quadrupoles which appear in DPS cross-section~(\ref{eq:XSec})
can be simplified as
\begin{align}
S_{4}\left(\boldsymbol{x}_{1},\boldsymbol{x}_{2},\boldsymbol{x}_{3},\boldsymbol{x}_{4}\right)-S_{2}\left(\boldsymbol{x}_{1},\boldsymbol{x}_{2}\right)S_{2}\left(\boldsymbol{x}_{3},\boldsymbol{x}_{4}\right) & \approx N\left(x,\,\boldsymbol{r}_{13},\,\boldsymbol{b}_{13}\right)+N\left(x,\,\boldsymbol{r}_{24},\,\boldsymbol{b}_{24}\right)-N\left(x,\,\boldsymbol{r}_{23},\,\boldsymbol{b}_{23}\right)-N\left(x,\,\boldsymbol{r}_{14},\,\boldsymbol{b}_{14}\right).
\end{align}

\begin{figure}
\includegraphics[width=9cm]{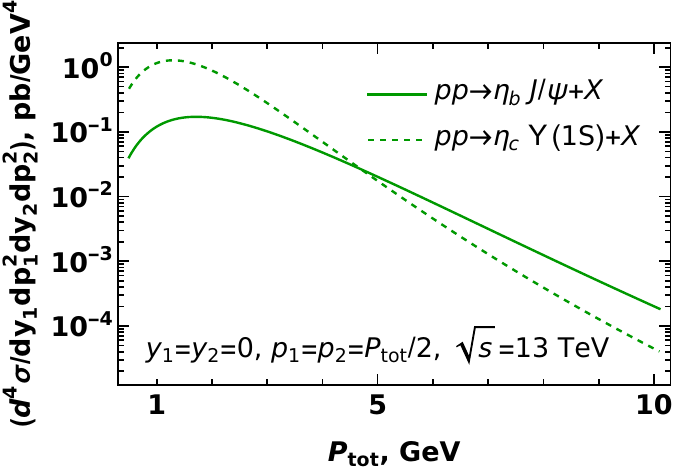}\includegraphics[width=9cm]{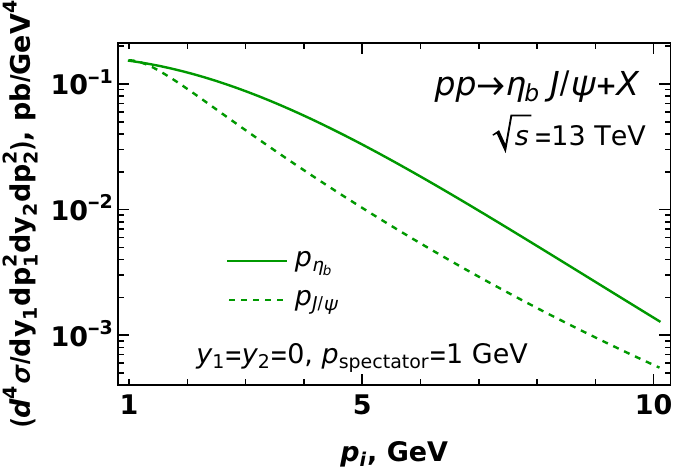}

\caption{\protect\label{fig:CGCBasic-pT}Left plot: The dependence of the cross-section
on the transverse momentum $P_{{\rm tot}}$ of the harmonium-bottomonium
pair assuming that it is shared equally between the final-state quarkonia.
Right plot: The dependence of the cross-section on the transverse
momentum of one of the quarkonia assuming that the other has fixed
transverse momentum $p=1$~GeV.}
\end{figure}

\begin{figure}
\includegraphics[width=6cm]{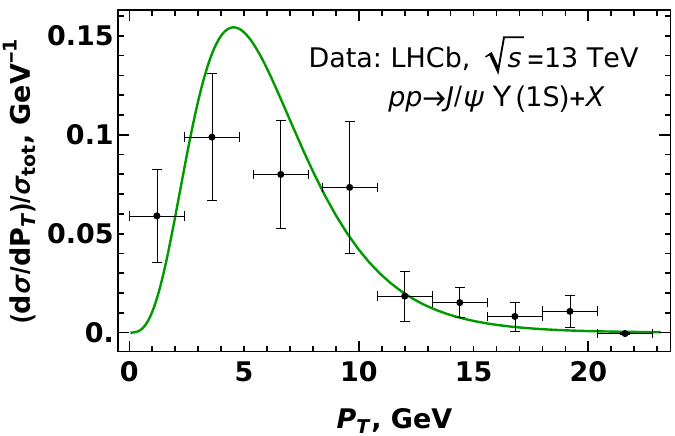}\includegraphics[width=6cm]{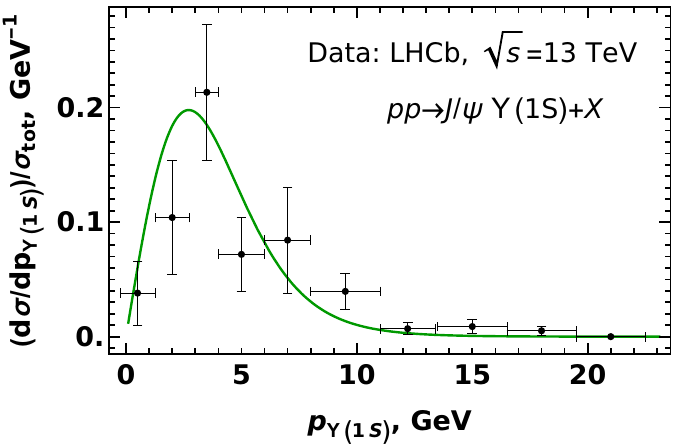}\includegraphics[width=6cm]{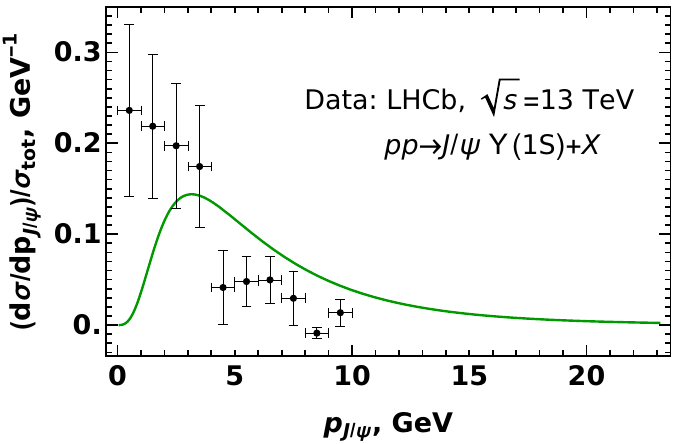}

\caption{\protect\label{fig:CGCBasic-LHCb}Comparison of the CGC predictions
(in dilute approximation) with experimental data from LHCb~\cite{LHCb:2023qgu}.
In the left plot we show the data for the cross-section as a function
of the total momentum of charmonia-bottomonia pair. The central and
right plots show the dependence on the transverse momentum of one
of the quarkonia, assuming that we integrated out the transverse momentum
of another. Note that in all plots we have shown the cross-section
normalized to $\sigma_{{\rm tot}}$, and the value of the latter is
different for the theoretical and experimental results (see the text
for more details).}
\end{figure}

The suppression of the dipole amplitude $N(x,\boldsymbol{r},\boldsymbol{b})$
at large impact parameters is controlled by the behavior of the factor
$T_{G}(b)$ defined in~(\ref{eq:gamma_eff}), and the effective cross-section
$\sigma_{{\rm eff}}$ which corresponds to transverse area of the
proton, can be estimated from
\[
\sigma_{{\rm eff}}\approx\pi\left\langle b^{2}\right\rangle =\pi\frac{\int d^{2}b\,b^{2}T_{G}(b)}{\int d^{2}b\,T_{G}(b)}\approx17.7\,{\rm mb},
\]
in agreement with other phenomenological estimates. In this way the
cross-section is fully determined by the dipole scattering amplitude
$N\left(\boldsymbol{r},\,\boldsymbol{b}\right)$. In the Figures~\ref{fig:CGCBasic-pT}
we have shown the cross-section as a function of the transverse momentum
of the produced quarkonia at central rapidities. We found that the
cross-section decreases rapidly as a function of $p_{T}$ of each
quarkonium. In the Figure~\ref{fig:CGCBasic-LHCb} we compare the
cross-section with recent experimental data from LHCb~\cite{LHCb:2023qgu}.
We found that overall the dilute approximation provides a reasonable
description of the \textit{shape} of the $p_{T}$-dependence. However,
the normalization of the cross-section, namely the value of 
\[
\sigma_{{\rm tot}}=\int dp_{T}\,\frac{d\sigma}{dp_{T}}
\]
 in dilute approximation is $\sigma_{{\rm CGC}}^{({\rm dilute})}\approx86\,{\rm pb}$
, which is $\sim$50\% smaller than the experimentally measured value
$\sigma_{{\rm exp}}^{({\rm LHCb})}=133\pm22\pm7\pm3\,{\rm pb}$. These
findings indicate that the suggested approach gives a fair qualitative
description of the data, though still there is a space for improvement
of the phenomenological parametrizations of the multipole matrix elements. 

\section{Conclusions}

\label{sec:Conclusions}In this study we applied the Color Glass Condensate
framework for analysis of the inclusive hadroproduction of heavy charmonia-bottomonia
pairs in the LHC kinematics. We found that there are several complementary
mechanisms which may be classified by number of gluons emitted from
projectile as Single- and Double Parton Scattering. We evaluated explicitly
the impact factors for the mechanisms which contribute in the leading
order in $\alpha_{s}$. For the DPS mechanism, the impact factors
in the large-$N_{c}$ limit factorize into a product of impact factors
of single quarkonia, so the DPS contribution can be reduced to the
well-known ``pocket formula''. The contribution of SPS is more sophisticated
and includes convolutions of impact factors with dipole, quadrupole,
sextupole and octupole scattering amplitudes (correlators of Wilson
lines). While there are some formal solutions parametrizations of
these objects based on solutions of JIMWLK equation, we could not
use them for phenomenological estimates of the cross-section due to
their complexity and excessive requirement of computational resources.
However, the phenomenological estimates of the cross-sections in the
dilute approximation suggest that our approach provides a fair qualitative
description of the LHCb data for $J/\psi+\Upsilon(1S)$ production. 

\section*{Acknowledgments}

We thank our colleagues at UTFSM university and Prof. Ian Balitsky
for encouraging and clarifying discussions. This research was partially
supported by Proyecto ANID PIA/APOYO AFB220004 (Chile) and Fondecyt
(Chile) grant 1220242. \textquotedbl Powered@NLHPC: This research
was partially supported by the supercomputing infrastructure of the
NLHPC (ECM-02)\textquotedbl .

 \end{document}